\documentclass[usenatbib]{article}
\usepackage{graphicx}
\usepackage{amssymb,amsmath}
\usepackage{color}

\newcommand{\be}{\begin{equation}}
\newcommand{\ee}{\end{equation}}
\newcommand{\lo}{{\cal L}_0}

\title{The Generalised Differential Image Motion Monitor}

\author{E. Aristidi,$^{1}$\footnote{E-mail:eric.aristidi@oca.eu}, A. Ziad$^{1}$, J. Chab\'{e},$^{2}$, Y. Fant\'ei-Caujolle$^{1}$,  C. Renaud$^{1}$, C. Giordano$^{1}$\\ 
$^{1}$Universit\'e C\^ote d'Azur, Observatoire de la C\^ote d'Azur, CNRS, Laboratoire Lagrange, France\\
$^{2}$Universit\'e C\^ote d'Azur, OCA, CNRS, IRD, G\'eoazur, 2130 route de l'Observatoire,\\ 06460 Caussols, France}

\date{\small MNRAS, 486, 915 (2019)}
\begin{document}

\maketitle

\label{firstpage}

\section*{abstract}
We present the Generalised Differential Image Motion Monitor. It is a compact instrument dedicated to measure 4 parameters of the optical turbulence: seeing, isoplanatic angle, coherence time and wavefront coherence outer scale. GDIMM is based on a small telescope (28cm diameter) equipped with a 3-holes mask at its entrance pupil. The instrument is fully automatic, and performs continuous monitoring of turbulence parameters at the Calern Observatory (France). This paper gives a description of the instrument, data processing and error budget. We present also statistics of  $3\frac 1 2$years of monitoring of turbulence parameters above the Calern Observatory.\\

{\sl \noindent keywords: 
Interferometers -- High Angular Resolution -- atmospheric effects -- site-testing.}
%
%
\section{Introduction}
\label{par:intro}
Atmospheric turbulence is responsible to the degradation of astronomical images observed through the atmosphere. Since the early 70's, many techniques have been developed to achieve diffraction limited resolution of observing instruments, namely speckle interferometry \cite{Labeyrie70}, long baseline interferometry \cite{Labeyrie75} and adaptive optics \cite{Rousset90}. Performances of these techniques rely on a good knowledge of atmospheric turbulence parameters, i.e. the seeing $\epsilon_0$, the isoplanatic angle $\theta_0$, the coherence time $\tau_0$ and the outer scale ${\cal L}_0$.

The 3 parameters $\epsilon_0$, $\theta_0$ and $\tau_0$ are of fundamental importance for adative optics (AO) correction: a large coherence time reduces the delay error, a small seeing value allows to close the loop  easily and benefit from a rather good correction, and a large isoplanatic angle reduces the anisoplanatic error, enlarges the sky coverage and allows very wide fields of correction (see \cite{Carbillet17} and references therein). {The outer scale ${\cal L}_0$ has a significant effect for large diameter telescopes (8m and above) and impacts low Zernike mode such as tip-tilt \cite{Winker91}.
}

Since several years, our group develops original techniques and instrumentation for measuring the optical turbulence of the atmosphere. Several prototypes were developped in the past, such as the generalized seeing monitor (GSM, \cite{Ziad00}) which has become a reference for monitoring the coherence parameters of the wavefront at ground level. In the last 15 years GSM was used in a large number of astronomical observatories and for prospecting potential new sites (see \cite{Ziad00} and references therein). 

The Generalized Differential Image Motion Monitor (GDIMM) was proposed in 2014 \cite{Aristidi14} to replace the aging GSM. It is a compact instrument very similar to a DIMM \cite{Sarazinroddier90}, with 3 sub-apertures of different diameters.  GDIMM observes bright single stars up to magnitude $V\sim 2$, at zenith distances up to 30$^\circ$, which is enough to ensure observability at any time/night of the year.

\begin{figure*}
 \parbox[c]{85mm}{
\includegraphics[width=8cm]{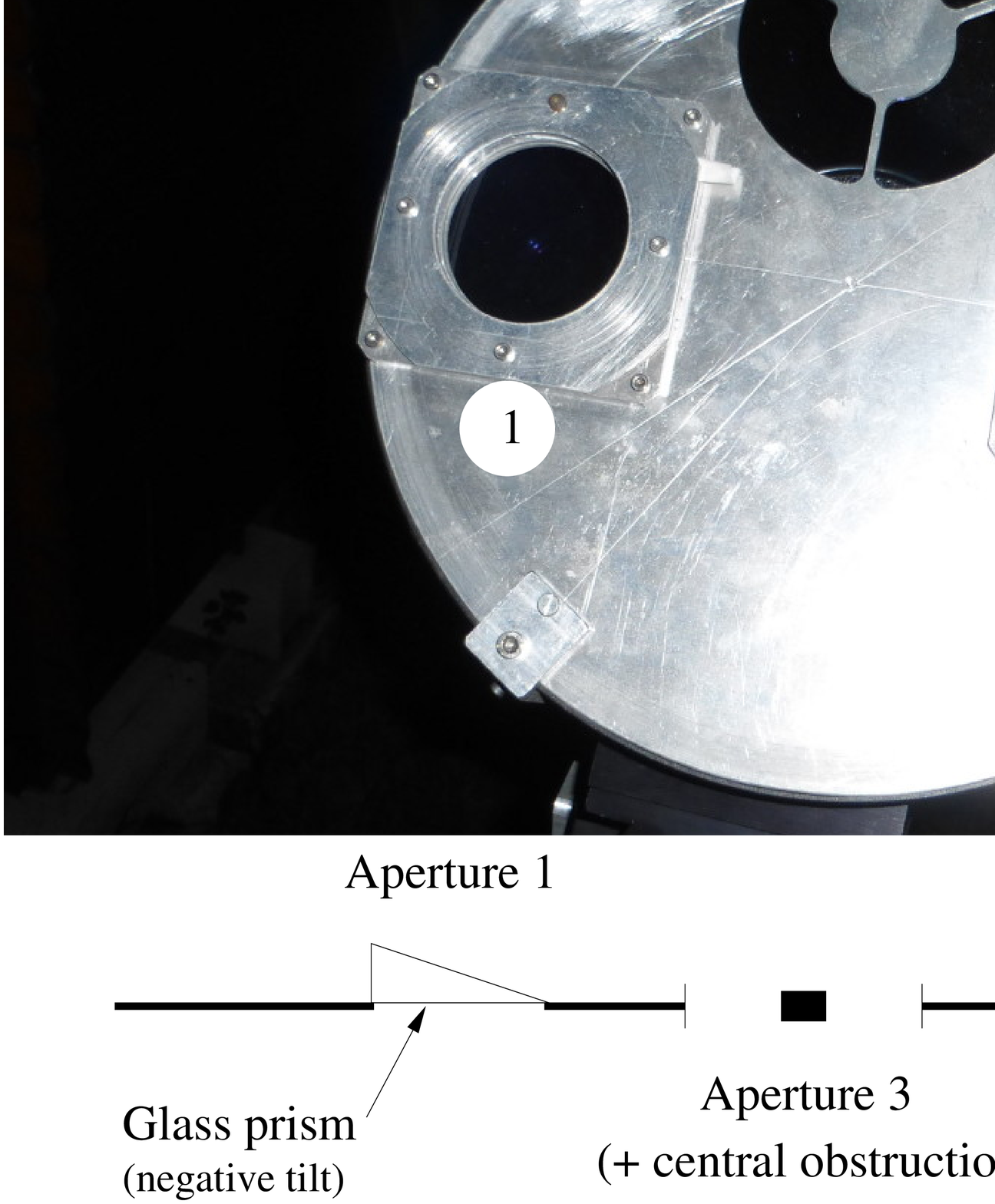}}\ 
\parbox[c]{85mm}{
\includegraphics[width=8cm]{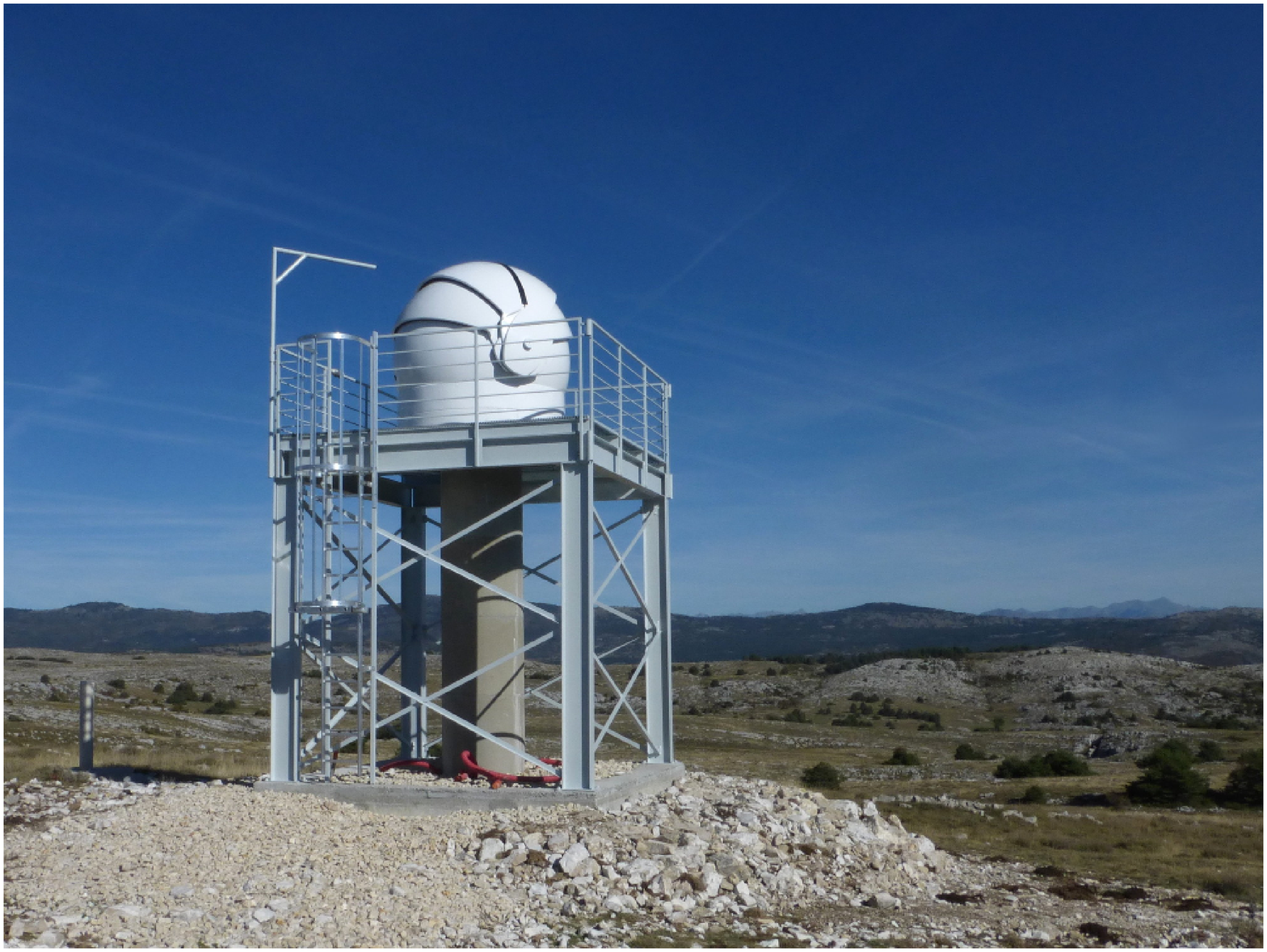}\\ \vskip 2.5mm \ \\}
\caption{Left: the pupil mask of GDIMM (bottom part is a sectionnal view). Right: the GDIMM dome on its 4m high tower at Calern Observatory.}
\label{fig:photogdimm}
\end{figure*}
After a period of developpement and tests in 2013--2015, the GDIMM is operational since the end of 2015, as a part of the Calern atmospheric Turbulence Station (C\^ote d'Azur Observatory -- Calern site, France, UAI code: 010, Latitude=$43^\circ 45' 13''$~N, Longitude=$06^\circ 55' 22''$~E). GDIMM provides continuous monitoring of 4 turbulence parameters ($\epsilon_0$,  $\theta_0$,  $\tau_0$ and  $\lo$)  above the Calern Observatory. Data are displayed in real time through a website ({\tt cats.oca.eu}), the idea being to provide a service available to all observers at Calern, as well as building a database to make long-term statistics of turbulence (before CATS, no such database existed for this site, despite his 40 years of activity as an astronomical site).
The other objective is that Calern becomes an operational on-sky test platform for the validation of new concepts and components in order to overcome current limitations of high angular resolution (HRA) existing systems. Several activities regarding adaptive optics are operated at the M\'eO \cite{Samain08} and C2PU \cite{Bendjoya12} telescopes and they benefit of the data given by the CATS station.

This paper is organised as follows: {Sect.~\ref{par:instrument} describes the instrument. Sect.~\ref{par:seeing} to~\ref{par:L0} present the method used to derive each  parameter (seeing, isoplanatic angle, coherence time and outer scale) and the associated error budget. }Sect.~\ref{par:results} is devoted to results obtained at the Calern observatory. A final discussion is presented
in Sect~\ref{par:conclusion}.

%
%
\section{Instrument description}
\label{par:instrument}
The GDIMM is based on a commercial Celestron C11 telescope (diameter 28cm), driven by an equatorial mount Astro-Physics AP900, controlled remotely by a computer. It is equipped with a pupil mask made of 3 sub-pupils (Fig.~\ref{fig:photogdimm}, left). Two sub-pupils are circular with a diameter $D_1=$6cm, separated by a distance $B=$20cm along the declination axis. Both are equipped with a glass prism oriented to give opposite tilts to the incident light. The third sub-aperture is circular, with diameter $D_3=$10cm and a central obstruction of 4cm and was designed to estimate the isoplanatic angle. It is protected by a glass parallel plate. A wide-field finder with a webcam is used to point stars and center them on the telescope.

The main camera is a Prosilica EC650. It offers a good sensitivity in the visible domain with a peak near the wavelength $\lambda=500$nm. The pixel size is 7.4$\mu$m. A Barlow lens enlarges the telescope focal to meet sampling requirements (we have $\lambda/D_1=7$ pixels and $\lambda/D_3=4$ pixels for $\lambda=500$nm). The camera allows short-exposure times and region-of-interest (ROI) definition to increase the frame rate. { An exposure time of a few milliseconds is required to observe stars of magnitude $V<2$ with sufficient SNR. The framerate is limited by the hardware, it is about 100 frames per second for our observations}. Such a high cadence is mandatory to properly sample the temporal variability of angles of arrival (AA) and to estimate the coherence time (see Sect.~\ref{par:calcultau0}).

\begin{figure}
\includegraphics[width=9cm]{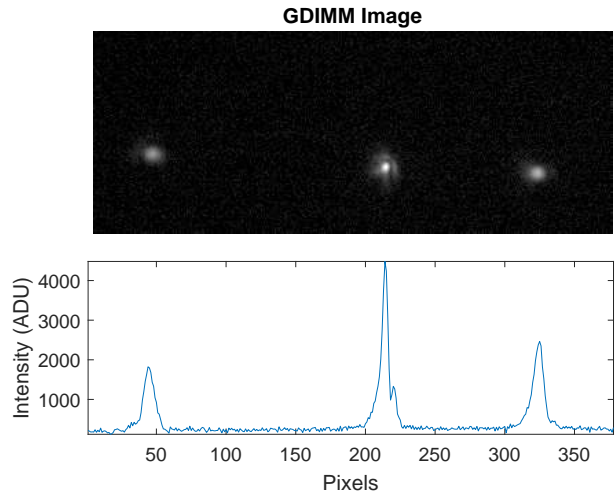}
\caption{Top: GDIMM instantaneous image, taken on 2018-04-04, 20:55UT on the star Regulus ($\alpha$~Leo) with an exposure time of 10ms. Bottom: 1D projection of the image (sum of lines).}
\label{fig:snapshot}
\end{figure}

An example of GDIMM short-exposure image is shown on Fig.~\ref{fig:snapshot}. It was obtained at Calern Observatory on April 4th, 2018 at 20h55UT on the star Regulus ($\alpha$~Leo, magnitude $V= 1.4$). The exposure time was 10ms {for this image}\color{black}. The central spot corresponds to the sub-pupil 3 (diameter 10cm); it is brighter than the two other ones, as expected. The first Airy ring is visible around the central spot: the seeing was $\epsilon_0=1.3$~arcsec for the wavelength $\lambda=500$nm (Fried diameter $r_0=8$cm, close to the pupil diameter). The image quality can be checked by computing the Strehl ratio of sub-images, using a simple formula proposed by \cite{Tokovinin02}. It is generally assumed that image quality is good when the Strehl ratio is over 30\% (this corresponds to phase distorsions lower than $\lambda/5$ over the pupil surface). For this example the 3 Strehl ratios are 0.79, 0.83 and 0.36 for spots corresponding to sub-pupils 1, 2 and 3.

The acquisition software is written in {\tt C++/QT}. It drives the whole observing sequence: dome opening, choice of the target star, telescope pointing, images acquisition, computation of turbulence parameters. The instrument is now fully automatic. It uses informations from a meteo station and a All-Sky camera to check observability. Observations are stopped if conditions degradate. 

The GDIMM is placed on the top of a 4m high concrete pillar, and protected by an all-sky dome (Fig.~\ref{fig:photogdimm}, right). A more detailed description is given in previous papers \cite{Aristidi14, Ziad17, Ziad18, Aristidi18}.

\subsection*{Data processing}
GDIMM data are based on sequences of two successive sets of $N=1024$ frames of a bright star, taken at exposure times $T=5$ms and $2T=10$ms. The full image size is 659$\times$493 pixels. Every frame contains 3 sub-images of the star, corresponding to the three sub-pupils of the instrument (see Fig~\ref{fig:snapshot}). Images are cropped in a rectangular zone (the ROI) of size 380$\times$150 pixels containing the 3 stellar spots. This allows to attain a cadence of 100 frames per second.  After sky background removal and thresholding, we calculate the three photocenters and integrated intensities. These raw data are logged into a file for optional further processing. A series of filters is then applied to control the data quality:
\begin{itemize}
\item Sub-image detection is made in 3 square boxes (size 30$\times$30 pixels for	lateral spots corresponding to pupils 1 and 2, 45$\times$45 pixels for the central spot, pupil 3) whose position is calculated on the first frame of the sequence. Sub-images for which the photocenter is too close to the box edge are rejected (this happens in case of strong wind or mount drift)
\item Outlier detection and rejection is made on photocenter coordinates and intensities.
\item Sub-images corresponding to the pupil 3 (diameter $D_3=10$cm) must be brighter than sub-images of pupils 1 and 2 (diameter $D_1=6$cm). They are rejected if it is not the case.
\item Drift correction is applied in by removing a linear trend on photocenter time series
\end{itemize}
The 4 turbulence parameters are then calculated (detailed description hereafter). The whole process (acquisition+processing) takes less than one minute of time. GDIMM provides one set of turbulence parameters every 2mn, they are sent to a database for real-time display on the CATS website ({\tt cats.oca.eu}). {Note that there is some dead time between two successive acquisitions to match this cadence of 2 minutes. We made this choice regarding the characteristic time of evolution of parameters, which is a few minutes (see \cite{Ziad16} and references therein). It we suppress the dead time, we can have a parameter quadruplet per minute with our current hardware. Some tests are currently be made to see if it can improve the parameter stability, especially for the outer scale estimation.
}

%
%
\section{Seeing measurements}
\label{par:seeing}
\subsection{Theory}
Seeing estimations by the GDIMM is based on differential motion. The principle of seeing estimation is well-known \cite{Sarazinroddier90}. It is based on variances of the photocenter difference of images produced by sub-pupils 1 and 2 (Fig.~\ref{fig:photogdimm}, left). The seeing $\epsilon_0$ (in radian) is computed using the following formulae~\cite{Tokovinin02}~:
\be
\epsilon_{0,l|t}=0.98 \,\left(\frac{D}{\lambda}\right)^{0.2}\:\left(\frac{\sigma_{l|t}^2}{K_{l|t}}\right)^{0.6}
\label{eq:seeing}
\ee
with
\begin{eqnarray}
K_l &=& 0.364\, (1-0.532 b^{-1/3})\nonumber \\ \ \label{eq:seeingK}\\
K_t &=& 0.364\, (1-0.798 b^{-1/3})\nonumber
\end{eqnarray}
where $B$ is the distance between the sub-apertures, $D$ their diameter, $b=B/D$, and $\lambda$ the wavelength, traditionnaly set to 500~nm as a standard. $\sigma_{l|t}^2$ are the longitudinal and transverse differential variances, calculated at the zenith (the correction is $\sigma^2(z=0)=\sigma^2(z)\, \cos(z)$ with $z$ the zenithal angle). Two estimations of the seeing are obtained for a given sequence, they are supposed to be the almost identical (isotropic hypothesis) and are averaged.
%
%

Differential variances (longitudinal and transverse) $\sigma^2_{l|t,T}$ and $\sigma^2_{l|t,2T}$ are calculated for sets corresponding to exposure times $T$ and $2T$. They  are compensated from the finite exposure $T$ time using an exponential interpolation as proposed by \cite{Tokovinin02}
\be
 \sigma^2_{l|t}=(\sigma^2_{l|t,T})^n \; (\sigma^2_{l|t,2T})^{1-n}
\label{eq:seeingcorrt}
\ee
This correction increase variances by a factor of the order of 10\% to 20\%. Two values of the seeing $\epsilon_{0,l|t}$ are deduced from Eq.~\ref{eq:seeing}, and averaged. 
%
%
\subsection{Error analysis}
\label{par:errorseeing}
\subsubsection{Statistical error.}
\label{par:staterr}
Variance of image motion at exposure times $T$ and $2T$ are computed from samples of $N=1024$
individual frames: they are then affected by statistical noise due to
the finite size of the sample. Assuming statistical independence
between the frames, the statistical error on the variance $\sigma^2$ (both at exposure times $T$ and $2T$)
is given by \cite{Frieden83}
\be
\frac{\delta \sigma^2}{\sigma^2}=\sqrt{\frac{2}{N-1}} 
\ee 
that propagates onto the seeing an error contribution $\delta \epsilon_0$.
With 1024 independent frames we have $\frac{\delta
\sigma^2}{\sigma^2}=4.4$\%. The error on the seeing is calculated from Eqs~\ref{eq:seeing} and~\ref{eq:seeingcorrt} and gives 
$\frac{\delta\epsilon_0}{\epsilon_0}\simeq 5$\%. This is the main source of uncertainty in our seeing estimations.
\subsubsection{Scale error}
Differential variances are obtained in units of pixel square and
require calibration of the pixel size. This is done by making
image sequences of binary star $\beta$ Cyg~AB (separation 34.6$''$).
We measured a pixel scale of $\xi=0.242\pm 0.003$$''$.

The uncertainty on $\xi$ propagates into
the differential variances when the conversion from pixels into
arcsec is performed. It gives a relative contribution on
the differential variances $\frac{\delta
\sigma^2}{\sigma^2}=0.6$\% and on the seeing
$\frac{\delta\epsilon_0}{\epsilon_0}=0.4$\%. 

The scale calibration has to be done regularly: the telescope tube is subject to thermal dilatations that result in slight variations $\delta F$ of the focal length $F$, especially during the transition between the summer and the winter. We measured relative variations $\frac{\delta F}F\lesssim 1\%$, leading to a relative uncertainty $\frac{\delta\epsilon_0}{\epsilon_0}\simeq 1$\% on the seeing. This remains lower than the statistical error.

\subsubsection{Background noise}
The sky background is an additive Poisson noise independent from the
stellar signal. Its influence on DIMM data is discussed in \cite{Tokovinin02}. 
 It biases the computed differential variances by a term 
\be \sigma_B^2=2
\frac{B^2}{I^2}\sum_{\mbox{\scriptsize window}} (x_{ij}-\bar x)^2
\label{eq:ron} 
\ee 
where $I$ is the total stellar flux, $B$ is the
sky background standard deviation and $x_{ij}$ the
coordinates of contributing pixels (the number of illuminated pixels in the star image
is typically of the order of 300  after thresholding and that defines the
``window'' over which the summation is made). With our data, the bias term is 
$\sigma_B^2\simeq 10^{-2}$ pixels$^2$, giving a relative error $\frac{\delta
\sigma^2}{\sigma^2}=0.2$\%. This is negligible compared to the statistical error.

Other instrumental noises include the readout noise of the CCD, and the error on the centroid determination. These errors were studied in details in the past (see \cite{Ziad94} and references therein) and have a very small contribution, orders of magnitude below the statistical error.

%
%
\section{Isoplanatic angle measurements}
\subsection{Theory}
The isoplanatic angle $\theta_0$ is estimated from the scintillation of a single star observed through the sub-pupil 3, with a diameter of 10~cm and a central obstruction of~4~cm (Fig.~\ref{fig:photogdimm}, left). The scintillation index is the ratio of the variance $\sigma_I^2$ of the stellar intensity, divided by the square of its mean value $\bar I$:
\be
\label{eq:scintindex}
s=\frac{\sigma_I^2}{\bar I^2}
\ee
The principle of the calculation is based on the similarity of the theoretical expressions of $\theta_0$ and the scintillation index $s$ \cite{Looshogge79, Ziad00}. $\theta_0$ is obtained (in arcsec) for a wavelength $\lambda=500$~nm by the following formula
\be
\theta_0^{-5/3}=A \, s
\label{eq:isop}
\ee
where $A=14.87$ is computed numerically from eqs. 19 and 21 of \cite{Ziad00}  using the value $h_0=10$km. The scintillation index $s$ is corrected from the zenithal distance $z$ by the formula $s(z=0)=s(z)\, \cos(z)^{8/3}$. 

Simultaneous measurements of the seeing and the isoplanatic angle make it possible to derive the equivalent turbulence altitude defined by \cite{Roddier82} as
\be
\bar{h}=0.31 \frac{r_0}{\theta_0} 
\label{eq:hmoy}
\ee
with $r_0=0.98 \frac{\lambda}{\epsilon_0}$ the Fried parameter. Statistics for $\bar h$ at Calern are presented in Section~\ref{par:results}. 
%
%
\subsection{Isoplanatic angle estimation}
\begin{figure}
\includegraphics[width=8cm]{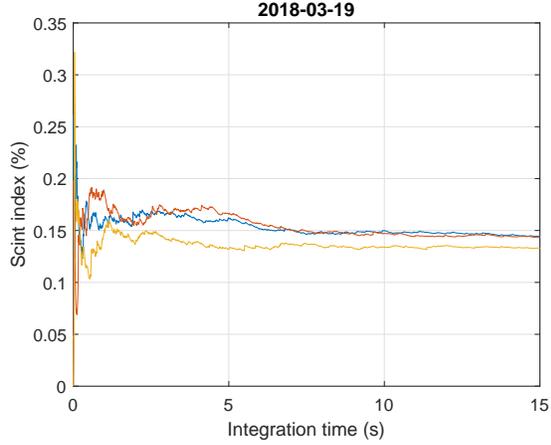} 
\caption{Scintillation index as a function of the integration time for 3 different data sets taken at Calern on the night of March 19$^{\rm th}$, 2018. The horizontal axis is limited to the range [0--15]s.}\label{fig:scintconv}
\end{figure}

Scintillation indexes (sub-image corresponding to pupil 3) $s_T$ and $s_{2T}$ are calculated for sets corresponding to exposure times $T$ and $2T$. 
{These sets are composed of $N=1024$ images, representing about 10s of data. This integration time appears to be long enough for the scintillation index to converge. To check that, we recorded long data sequences (up to 4000) images and calculated the scintillation index for integration times varying from 0 to 40s. The result is shown in Fig.~\ref{fig:scintconv} for 3 different sets taken at Calern on the night of March 19$^{\rm th}$, 2018. Scintillation indexes show satisfactory convergence (below 2\%) after 10s of integration time.
}

Compensation from the finite exposure time is made by linear extrapolation on scintillation indexes as proposed by \cite{Ziad00}
\be
s=2 s_T-s_{2T}
\label{eq:isopcorrt}
\ee
This compensation is more critical on the scintillation than on the differential variances. The correction can be of the order of 30\%--50\%. The isoplanatic angle is then derived from Eq.~\ref{eq:isop}.
%
%
%
\subsection{Error analysis}
\subsubsection{Statistical error}
The isoplanatic angle is estimated from the scintillation index $s$ via Eq.~\ref{eq:scintindex}. { Two estimates $s_T$ and $s_{2T}$ are made, corresponding to exposure times $T$ and $2T$, and combined to obtain the scintillation index corrected from the exposure time effects (Eq.~\ref{eq:isopcorrt}). The error on $s_T$ (same as $s_{2T}$) is}
\be 
\frac{\delta s_T}{s_T}=\frac{\delta \sigma_I^2}{\sigma_I^2}+2 \frac{\delta\bar{I}}{\bar{I}}
\label{eq:statnoiseisop}
\ee
If we assume statistical independence between frames, the first term is the same as for the seeing and the second is $\frac 2{\sqrt N}$. We get
\be
\frac{\delta s_T}{s_T}=\sqrt{\frac{2}{N-1}} + \frac 2{\sqrt N}\simeq 10\%
\label{eq:err_s}
\ee
{
However in case of slow wind speed in the upper atmosphere (the major contributor to scintillation) the number of independent frames within an image cube is reduced to a number $N_I<N$ and we must replace $N$ by $N_I$ in the previous equation. An order of magnitude of $N_I$ is given by the ratio
\be
N_I=\frac{N t_e}{D_3/v}
\ee
where $D_3$ is the diameter of the sub-pupil 3, $N t_e$ the integration time (10s at a framerate of 100~Hz) and $v$ the wind speed of atmospheric layers contributing to the scintillation (high altitude layers). We do not know $v$, but we can have its order of magnitude by looking at the distribution of the effective wind speed $\bar v$ defined in Eq.~\ref{eq:tau0}. At Calern observatory, the distribution is bimodal (as shown in Fig.~\ref{fig:histoveffsum}) and high layers have a speed of the order of 13m/s. Taking this value for $v$, we obtain $N_I\simeq 1300$, which is the same order of magnitude as the number $N$ of frames in a data cube.
Using Eq.~\ref{eq:isopcorrt}, we obtain the relative statistical error on the zero exposure time scintillation index
\be
\frac{\delta s}{s}\simeq 15\%
\ee

There is another error source depending on the constant $A$ in Eq~\ref{eq:isop}. $A$ is indeed a function of an altitude parameter $h_0$ defined in eq.~21 of \cite{Ziad00}. The relation $A(h_0)$ is a function of the pupil geometry and is analytic. Its dependence with $h_0$ remains weak, we found that the relative error  
\be
\frac{\delta A}{A} \lesssim 5\%
\ee
in the range $h_0 \in [1, 25]$km. The relative statistical error on the isoplanatic angle is 
\be
\frac{\delta \theta_0}{\theta_0}=\frac 3 5 \frac{\delta A}{A}+ \frac 3 5 \frac{\delta s}{s} \simeq 15\%
\ee
} 
\subsubsection{Sky background}
The presence of a sky background on individual images introduces a bias on the estimation of the mean stellar
intensity $\bar{I}$, its
standard deviation $\sigma_I$ and then on the scintillation index $s$. The observed background on our images is typically 40 ADU/pixel. Its relative contribution to the stellar flux (integrated on the star image) represents about 30\% for bright stars such as Deneb ($\alpha$ Cyg, magnitude $V=1.2$) observed by the GDIMM. To estimate the bias, let us introduce the following variables:
\begin{itemize}
\item $B$, the background intensity collected over the $N_I$ pixels illuminated by the star after threshold application, $\bar{B}$ and $\sigma^2_B$ its mean and variance. $B$ is a Poisson random variable, it must verify $\sigma_B=\sqrt{\bar B}$, that was well verified on images.
\item $I_t$ the total intensity (background+stellar flux) collected over the $N_I$ pixels.
\end{itemize}
The stellar flux is given by $I=I_t-B$, the measure being $I_t$. The mean $\bar{I}$ is biased by the term $\bar{B}$. This bias is estimated  and removed as indicated above, but the background fluctuations lead to an error $\delta I$ on the estimation of $\bar{I}$ equal to $\delta I=\sigma_B\simeq \sqrt{\bar B}$. Similarly, the intensity variance $\sigma_I^2$ is biased by a term $\sigma_B^2$. 

The error on the scintillation index is calculated from Eq.~\ref{eq:statnoiseisop} taking $\delta \sigma_I^2=\sigma_B^2$ (bias on intensity variance) and $\delta\bar{I}=\sigma_B$. Typical values are, in ADU units: $\sigma_B\simeq 240$, $\bar{I}\simeq 100000$, $\sigma_I\simeq 30000$. That gives a background error on the scintillation index $\frac{\delta s}{s}\le 1\%$, which is an order of magnitude below the statistical error.

%
%
\section{Coherence time measurements}
\label{ctime}
\subsection{Theory}
The coherence time $\tau_0$ relevant for AO applications, is defined by \cite{Roddier81}
\be
\tau_0=0.31\frac{r_0}{\bar v}
\label{eq:tau0}
\ee
where $\bar v$, the effective wind speed, is a weighted average of the wind speed on the whole atmosphere. It can be estimated \cite{Ziad12, Aristidi14, Ziad17} from the temporal structure functions $D_{x|y} (\tau)$ of the AA in the $x$ (resp. $y$) direction (parallel to the declination (resp. right ascension)).  This function is zero for $\tau = 0$ and saturates to a value $D_s$ for large $\tau$, and its characteristic time 
\be
D(\tau_{AA,x|y})=\frac{D_s}{e}
\label{eq:tauaa}
\ee
defines the decorrelation time of AA fluctuations in directions $x$ and $y$. To calculate the effective wind speed $\bar{v}$, we make use of the work by \cite{Conan00} and \cite{Ziad12} who gave two approximations of $\bar{v}$ (in m/s) corresponding to two different regimes:
\begin{itemize}
\item For $\tau_{AA,x|y} > \frac D{\bar v}$ 
\be
\bar v=10^3 D\, G^{-3} \left[\tau_{AA,x}^{\frac 1 3} + \tau_{AA,y}^{\frac 1 3}  \right]^{-3}
\label{eq:veff}
\ee
where $D$ is the sub-pupil diameter and $G$ a constant \cite{Conan00}:
\be
G=\frac{(1-e^{-1}) (3.001 K^{\frac 1 3} + 1.286 K^{\frac 7 3})   +   e^{-1} (2.882+1.628 K^2)}{0.411+0.188 K^2}
\ee
with $K=\frac{\pi D}{\lo}$. This case is met almost all  the time with small pupils as GDIMM ones.
\item For $\tau_{AA,x|y} < \frac D{\bar v}$ (this case was never observed with our data):
\be
\bar v=\frac{D\, \sqrt{G_1}}2 \left[\tau_{AA,x}^{-2} + \tau_{AA,y}^{-2}  \right]^{\frac12}
\ee
with
\be
G_1=\frac{2.62}{e} \, \left(1-1.04 K^{\frac 1 3} +0.57 K^2 - 0.45 K^{7/3} 
\right)
\ee
\end{itemize}
We obtain 3 values of $\bar v$ for the 3 sub-pupils, which are averaged. The coherence time $\tau_0$ is eventually calculated from $r_0$ and $\bar v$ using Eq.\ref{eq:tau0}.
%
%
\subsection{Coherence time estimation}
\label{par:calcultau0}
\begin{figure*}
\begin{center}
\includegraphics[width=8cm]{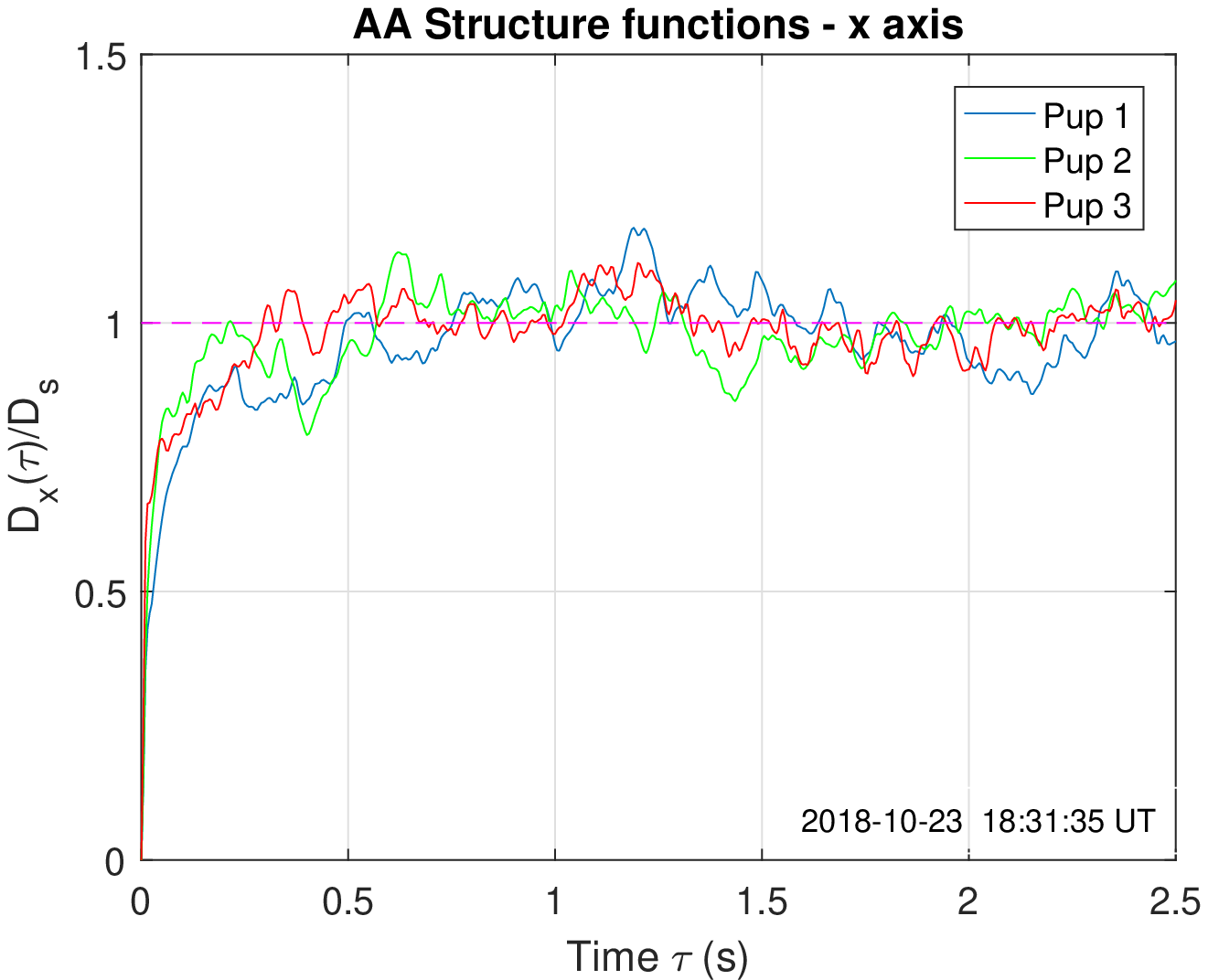}
\includegraphics[width=8cm]{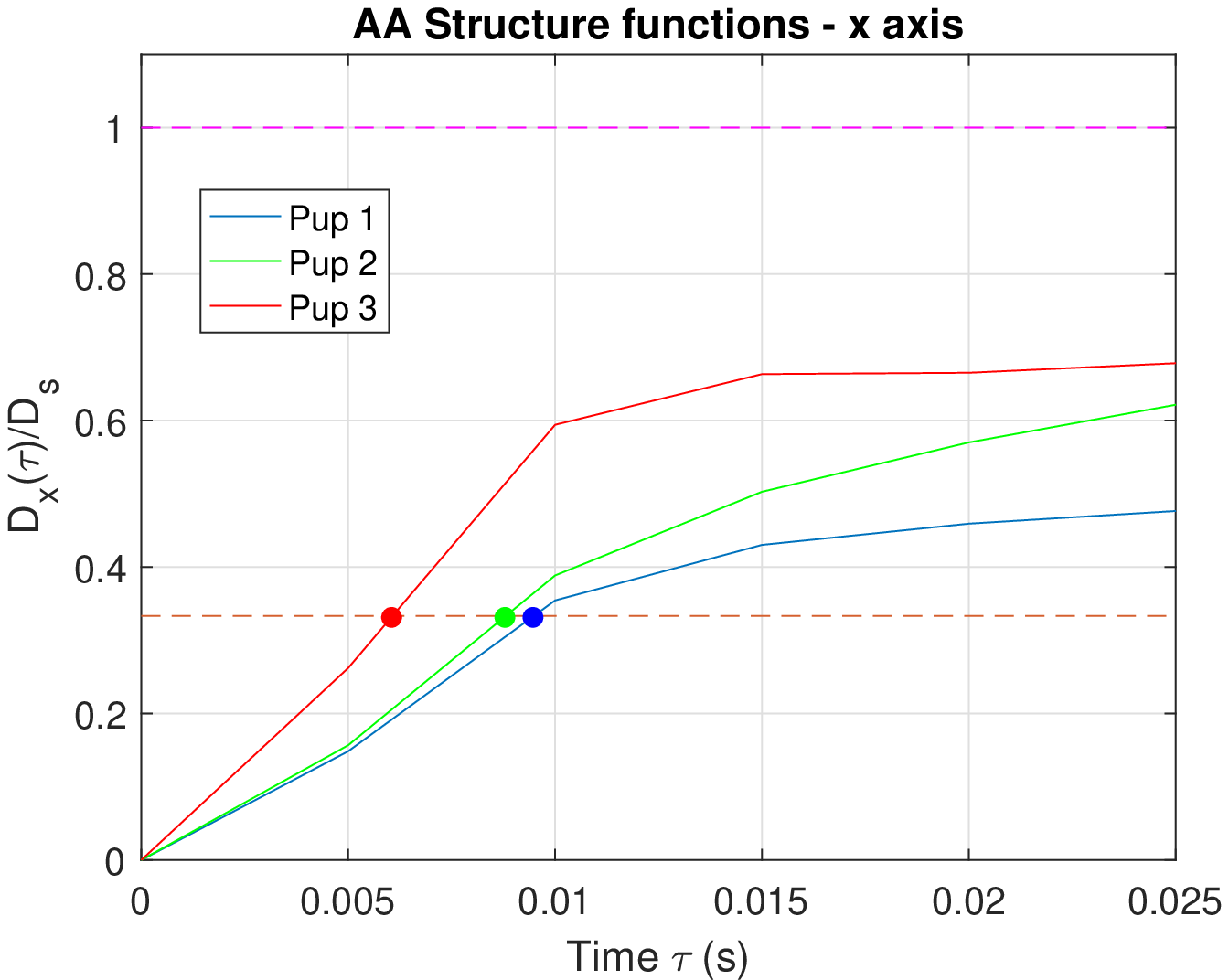}
\caption{Example of normalised structure functions of AA fluctuations along the $x$ axis, calculated for the 3 sub-pupils (and compensated from exposure time). Left: structure functions $\frac{D_x(\tau)}{D_s}$ divided by their saturation value. Right: zoom for $\tau\in[0, 25]$ms. The 3 curves intersect the line $\frac{D_x(\tau)}{D_s}=\frac 1 e$ (brown dashed line) at $\tau=\tau_{AA}$ (circles).}
   \label{fig:structfn}
\end{center}
\end{figure*}
We remarked that the framerate (100 frames/second) is slightly variable: the first operation is then to resample time series of photocenter coordinates with a constant time step $\delta t$ (after some trials, we chose $\delta t=5$~ms). 12 structure functions $D(\tau)$ are computed for the 12 photocenter series (2 coordinates for 3 sub-images, and two frames sets for exposure times $T$ and $2T$) using the direct expression
\be
D_{x|y}(\tau)=\langle \left[x|y(t) - x|y(t+\tau)\right]^2\rangle
\ee
where $\langle \rangle$ stands for ensemble average over the $N=1024$ frames. Structure functions are compensated from finite exposure time using the same method as for the seeing:
\be
D_{x|y} (\tau)=D_{x|y,T} (\tau)^{n} \; D_{x|y,2T} (\tau)^{1-n}
\ee
where $D_{x|y,T}$ and $D_{x|y,2T}$ are calculated on image cubes taken with exposure times of $T$ and $2T$, and $n=1.75$. An example of structure functions is shown in Fig.~\ref{fig:structfn}. Curves correspond to the $x$ axis (declination) and were divided by their respective saturation value $D_s$. One can remark that the saturation is attained after 0.3--0.4s, and that there are some fluctuations of $D_x(\tau)$ in the saturation regime. These fluctuations are the main source of uncertainty on the determination of $\tau_{AA}$, as discussed in Sect.~\ref{par:errorstau}. The graph on the right is a zoom for small values of $\tau$: curves intersect with the line $\frac{D_x(\tau)}{D_s}=\frac 1 e$ at $\tau_{AA,1}=9.5$ms, $\tau_{AA,2}=8.8$ms and  $\tau_{AA,3}=6.1$ms.

For each sub-pupil, the effective wind speed $\bar v$ is calculated from Eq.~\ref{eq:veff}. The three values of  $\bar v$ are then averaged. 

%
%
\subsection{Error analysis}
\label{par:errorstau}
The coherence time is deduced from the AA decorrelation time $\tau_{AA}$ defined by Eq.~\ref{eq:tauaa}. To calculate the error on $\tau_{AA}$, we express the finite difference
\be
\delta D_0(\tau) \simeq D_0'(\tau)\: \delta \tau
\ee 
where $D_0(\tau)=\frac{D(\tau)}{e}$ is the normalised structure function and $D_0'(\tau)$ the derivative of $D_0$. Then, it is possible to estimate the error $\delta\tau$ at $\tau=\tau_{AA}$:
\be
\delta \tau=\frac{\Delta D_0}{D_0'(\tau_{AA})}
\ee
The error on $D_0$ can be estimated as the standard deviation of the structure function in the saturation zone, typical values are 10\% to 20\%. The derivative $D_0'(\tau_{AA})$ can be estimated by the slope of the structure function at $\tau=\tau_{AA}$. Errors on on $\tau_{AA}$ were calculated for each of the 6 structure function, for a 3~months data sample. We found a typical error on $\sim 30\%$. Relative errors on $\tau_{AA,x|y}$ for each sub-pupil and are summarised in the tabular below

\begin{tabular}{l|c|c|c|c|c|c}\hline
&\multicolumn{2}{c}{sub-pup. 1} & 
\multicolumn{2}{c}{sub-pup. 2} &
\multicolumn{2}{c}{sub-pup. 3} \\ 
&$x$ & $y$ & $x$ & $y$ & $x$ & $y$\\ \hline
$\frac{\Delta \tau}{\tau_{AA}}$ &29\%  &  35\%  &  29\%  & 37\%  & 27\%  & 27\% \\ \hline
\end{tabular}
{

The error on $\tau_{AA}$ propagates to the effective wind speed, giving a contribution $\delta_{v,\tau}$ to the uncertainty on $\bar v$, obtained by differentiation of Eq.~\ref{eq:veff}. For a relative error of $30\%$ on $\tau_{AA}$, this contribution $\delta_{v,\tau}$ is of 10\% (for $\tau_{AA}=6$ms) to 20\% (for $\tau_{AA}=24$ms).

In addition, the effective wind speed $\bar v$ calculated from Eq.~\ref{eq:veff} needs an estimate of the outer scale ${\cal L}_0$. However, at discussed in section~\ref{par:errorsL0}, the outer scale is strongly filtered and a measurement is not always available. In this case the standart value ${\cal L}_0=20$m is used. This results in a bias $\delta_{v,L}$ on the effective wind speed. This bias remains below 20\% for outer scales ${\cal L}_0 \in [10, 40]$m, which covers the majority of the situations on traditional sites. 
Combining these two contributions, the relative uncertainty on $\bar v$ is then $\frac{\delta_v}{\bar v}\simeq 20$\% to 30\%.
}

The error $\delta_{\tau 0}$ on the coherence time $\tau_0$ is obtained from Eq.~\ref{eq:tau0}:
\be
\frac{\delta_{\tau 0}}{\tau_0}=\frac{\delta_{\epsilon_0}}{\epsilon_0}+ \frac{\delta_v}{\bar v} \simeq 25\% \; \mbox{ to } \; 35\%
\ee

%
%
\section{Outer scale measurements}
\label{par:L0}
\subsection{Theory}
The outer scale is, among the 4 turbulence parameters measured by GDIMM, the most difficult to estimate with a small instrument. In previous papers \cite{Ziad94, Aristidi14} we proposed to make use of variances of the absolute motions of sub-images to estimate the outer scale $\lo$. These absolute variances (in square radians) are given by \cite{Ziad94}
\be
\label{eq:varabs}
\sigma_D^2=0.17\, \lambda^2 r_0^{-5/3}\, (D^{-1/3}-1.525 \lo^{-1/3})
\ee
\begin{figure}
\begin{center}
\includegraphics[width=9cm]{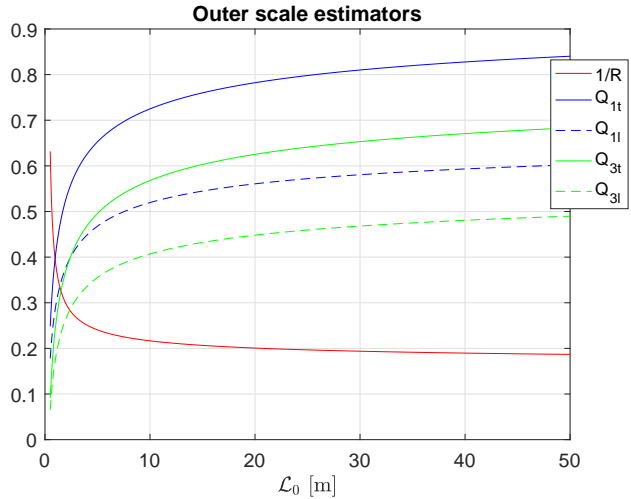}
\caption{Outer scale estimators: $1/R$ (Eq.~\ref{eq:ratioRL0}) and $Q_{i,l|t}$ (Eq.~\ref{eq:ratioQL0}) as a function of $\lo$.}
   \label{fig:diffvarl0}
\end{center}
\end{figure}
Because of telescope vibrations, direct estimation of $\lo$ from absolute variances using Eq.~\ref{eq:varabs} is not reliable. Our first idea, following the work by~\cite{Ziad94}, was to use the inverse relative difference of variances measured with sub-pupils 1 (or 2) and 3 (diameters 6cm and 10cm), i.e. 
\be
R=\frac{\sigma_{D1}^2}{\sigma_{D1}^2 - \sigma_{D3}^2}=\frac{D_1^{-1/3}-1.525 \lo^{-1/3}}{D_1^{-1/3}-D_3^{-1/3}}
\label{eq:ratioRL0}
\ee
But with our values for $D_1$ and $D_3$, the variation is weak for decametric values of $\lo$, as illustrated by Fig.~\ref{fig:diffvarl0} and Table~\ref{table:estl0}. We have $1/R=0.216$ for $\lo$=10m and 0.200 for $\lo$=20m. To extract reliable $\lo$ from this estimator, we need high precision on variances (about 1\%), which is not the case (the statistical error on variances is of the order of 5\% as discussed in Sect.~\ref{par:errorseeing}, and there is some bias from telescope vibration).

We then looked for another estimator for $\lo$, and found that it was possible to use  the ratio of absolute to differential variances of image motion:
\be
\label{eq:ratioQ}
Q_i=\frac{\sigma_{Di}^2}{\sigma_{l|t}^2}
\ee
where $\sigma_{D_i}^2$ is the absolute variance corresponding to the sub-pupil $i$, and $\sigma_{l|t}^2$ the longitudinal or transverse differential variance used to calculate the seeing (Eq.~\ref{eq:seeing}).  This gives two expressions for the ratios $Q_i$
\be
\begin{array}{lll}
Q_{i,t} & = & \displaystyle \frac{\sigma_{D_i}^2}{\sigma_{t}^2}\; = \; 0.17 \frac{D_i^{-1/3}-1.525 \lo^{-1/3}}{0.364 D_1^{-1/3} -0.2905 B^{-1/3}}\\ \\

Q_{i,l} & = & \displaystyle\frac{\sigma_{D_i}^2}{\sigma_{l}^2}\; = \; 0.17 \frac{D_i^{-1/3}-1.525 \lo^{-1/3}}{0.364 D_1^{-1/3} -0.1904 B^{-1/3}}
\end{array}
\label{eq:ratioQL0}
\ee
Using absolute variances from the 3 sub-pupils, we get 6 estimations of $\lo$, from which we take the median value. Note that the absolute variance at the numerator of Eq.~\ref{eq:ratioQ} may be contaminated by telescope vibrations. Hence we use only the $x$ direction (declination axis) to compute absolute variances to reduce oscillations from the motor of the mount. Fig.~\ref{fig:diffvarl0} shows the variation of ratios $1/R$ and $Q_{i,l|t}$ as a function of $\lo$. All estimators have weak dependence with decametric $\lo$, but the ratios $Q_i$ are a little more sensitive. In Table~\ref{table:estl0} we computed the expected $Q_i$ ratios for $\lo=10$m and $\lo=20$m, and the required precision on variances to discriminate between the 2 values of $\lo$. We found that this required precision is 4 to 5\% for the ratios $Q_i$, while it was 1\% for the ratio $R$. 

\begin{table}
\begin{tabular}{c|ccc}\\ \hline
 & $\lo=$ 10m & $\lo=20$m &  Required precision \\
 &            &           &  on variances \\ \hline
$1/R$     & 0.216   & 0.200     & 1\% \\
$Q_{1,l}$ & 0.520   & 0.560     & 4\% \\
$Q_{1,t}$ & 0.725   & 0.782     & 4\% \\
$Q_{3,l}$ & 0.407   & 0.448     & 5\% \\
$Q_{3,t}$ & 0.578   & 0.625     & 5\% \\ \hline
\end{tabular}
\caption{Value of ratios $1/R$ (Eq.~\ref{eq:ratioRL0}) and $Q_{i,l|t}$ (Eq. \ref{eq:ratioQL0}) for $\lo=10$m and $\lo=20$m. Column 4 is the required precision on variances to discriminate between the 2 values of $\lo$.}
\label{table:estl0}
\end{table}
Note that this estimator uses ratios of variances and is therefore independent of scale calibration. Also,  we can remark that it is not necessary to have pupils of different diameters, the method should work with any DIMM or with a Shack-Hartmann (however, in this case it will not be possible to filter data with $H$ invariants presented hereafter).\\
\subsubsection*{$H$ Invariants}
\label{par:Hinv}
Combining Eqs \ref{eq:seeing} and \ref{eq:varabs}, we calculated the following ratios
\be
\begin{array}{lll}
H_{t} & = & \displaystyle \frac{\sigma_{D_i}^2-\sigma_{D_3}^2}{\sigma_{t}^2}\; = \; \frac{ 0.17 (D_1^{-1/3}-D_3^{-1/3})}{0.364 D_1^{-1/3} -0.2905 B^{-1/3}}\\ \\

H_{l} & = & \displaystyle\frac{\sigma_{D_i}^2-\sigma_{D_3}^2}{\sigma_{l}^2}\; = \; \frac{0.17 (D_1^{-1/3}-D_3^{-1/3})}{0.364 D_1^{-1/3} -0.1904 B^{-1/3}}
\end{array}
\label{eq:Hinv}
\ee
Where $i=1,2$ refers to sub-pupil 1 or 2 (they have the same diameter $D_1=D_2=6$cm). These ratios appear, at the first order, to be independent of turbulence conditions, so we named them ``$H$ invariants''. In fact this invariance is valid for large outer scales ($\frac{\lo}{D_i}\gg 1$). There is indeed a weak dependence of differential variances $\sigma^2_{l|t}$ with the outer scale \cite{ZiadThese}. This dependence is generally omitted in seeing estimations (Eq.~\ref{eq:seeing} and~\ref{eq:seeingK}). It can be estimated using eqs.~5.4 and~5.8 of~\cite{Conan00}. For pupils of diameter of 6cm, the effet of the outer scale on differential variances is under 0.1\% for $\lo>10$m and over 3\% for $\lo<1$m. The impact on $H$ invariants is $\lesssim 0.03$\% for $\lo>10$m and becomes greater than 2\% for $\lo<1$m (these very low outer scales are nevertheless exceptional: at Calern they correspond to less than 0.5\% of measured values).\\
Values of $H$ corresponding to our instrument are 
\be
H_t=0.1567 \quad \mbox{and} \quad H_l=0.1128
\ee
These invariants are easy to calculate and can be used as a filter to reject bad data (contaminated by telescope vibrations). More discussion will be presented in Sect.~\ref{par:dataprocL0}.
%
%
\subsection{Outer scale estimation}
\label{par:dataprocL0}
Estimation of the outer scale requires absolute variances $\sigma_{Di}^2$ of AA fluctuations for each pupil (in the $x$ direction only). { As for differential variances used for seeing estimation, absolute variances are calculated from each image cube, and corrected from exposure time, following the same process as for differential variances (Eq.~\ref{eq:seeingcorrt}). One obtains a set of absolute and differential variances every 2mn.}
To reduce noise, time series of variances (both absolute and differential) are smoothed by a temporal sliding average. After some trials, the width of the temporal window was set to 10mn, leading to an average of 5 successive variances, reducing the error by a factor $\sqrt 5$ (see Sect.~\ref{par:errorsL0}). 

\begin{figure}
\begin{center}
\includegraphics[width=8cm]{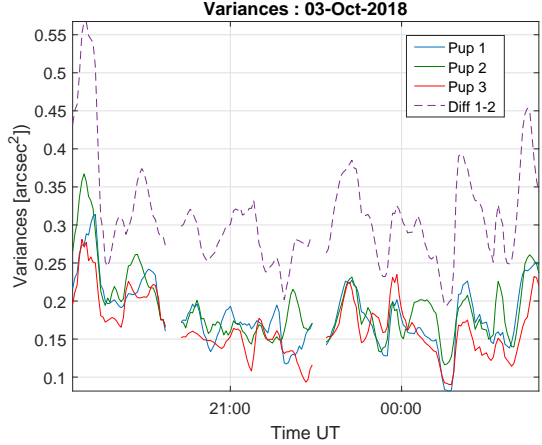}
\caption{Time series of variances observed at Calern on 2018-10-03. Solid lines: absolute variance in the $x$ direction (declination) for the 3 subpupils. Dashed line: differential longitudinal variance $\sigma_l^2$ between pupils 1 and 2. These variances were smoothed by a 10mn large sliding average.}
   \label{fig:varts}
\end{center}
\end{figure}

Fig.~\ref{fig:varts} shows an example of the evolution of these smoothed variances for the night of  2018-10-03. Two things can be notices on these curves:
\begin{itemize}
\item The variance $\sigma_{D3}^2$ corresponding to the sub-pupil 3 should be smallest than $\sigma_{D1}^2$ according to Eq.~\ref{eq:varabs}. This is not always the case, fluctuations are sometimes larger than the expected difference.
\item The differential variance $\sigma_l$ between sub-pupils 1 and 2 is almost two times greater than absolute variances. This is good news: it means that the AA fluctuation signal is not dominated by correlated vibrations due to the telescope mount.
\end{itemize}

\begin{figure*}
\begin{center}
\includegraphics[width=8cm]{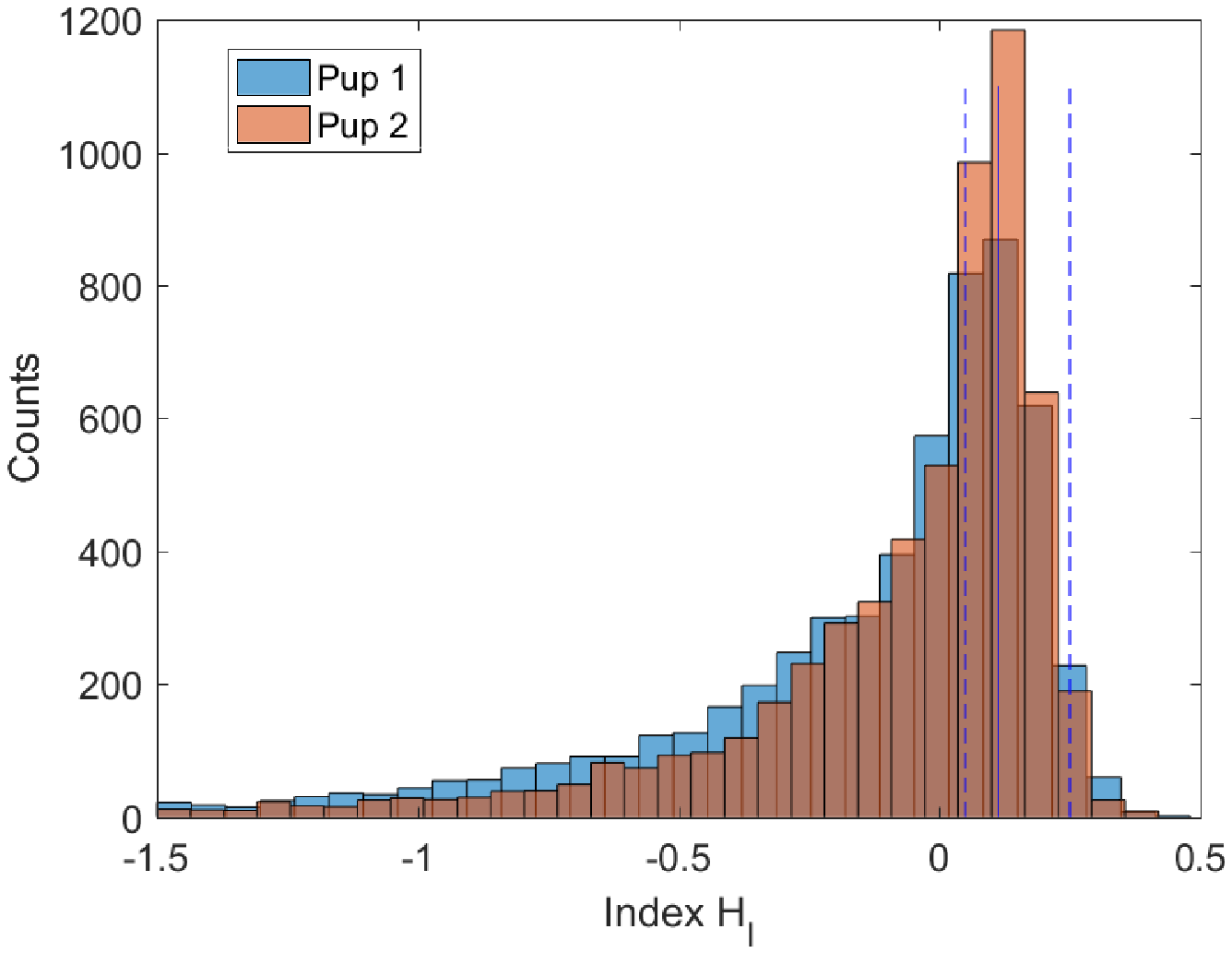}
\includegraphics[width=8cm]{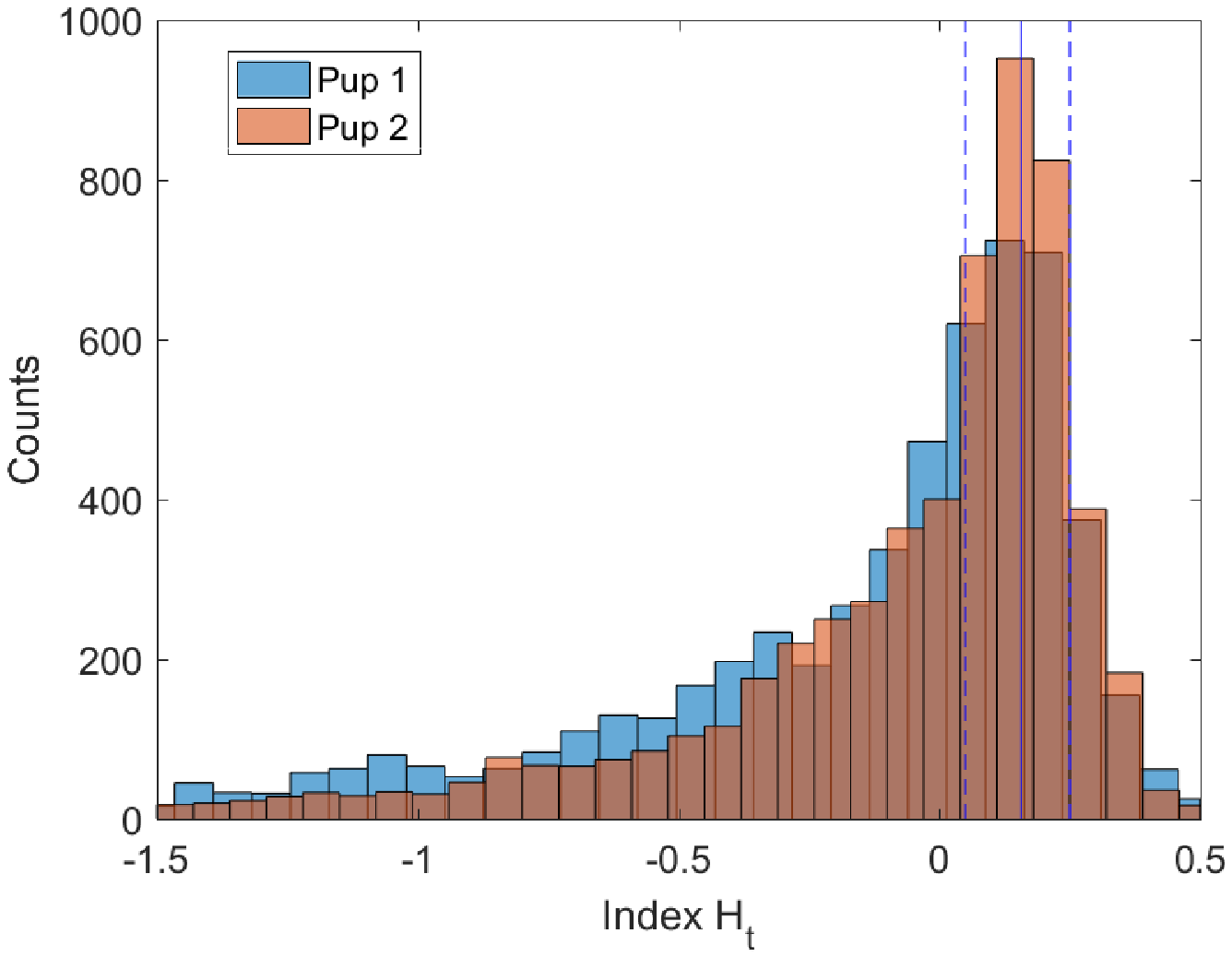}
\caption{Histograms of invariants $H_l$ (left) and $H_t$ (right) for measured at Calern during the period August--October 2018. Blue (esp. orange) bars correspond to sub-pupil 1 (resp. 2). The vertical solid line is the theoretical value, and the two dashed lines are rejection thresholds.}
   \label{fig:histoH}
\end{center}
\end{figure*}

The 6 ratios $Q$ are calculated from Eq.~\ref{eq:ratioQ} leading to 6 estimations ${\cal L}_{0,i}$ of the outer scale. Then, we calculate invariants $H_{l|t}$ (Eq.~\ref{eq:Hinv}) to be used as a filter for bad data. Histograms of $H$ invariants obtained during a 3~month period (August--October 2018) are displayed in Fig.~\ref{fig:histoH}. They present a peak for the theoretical value ($H_t=0.1567 $ and $H_l=0.1128$), and somewhat large dispersion around it. This dispersion result mainly from contamination of variances by noise and/or telescope vibrations (there is also a weak contribution due to the dependence of $H_{l|t}$ with the outer scale). After some trials, we decided to reject data for which $H_{l|t}>0.25$ or $H_{l|t}< 0.05$. That led to rejection of about 70\% of the individual outer scales ${\cal L}_{0,i}$. The final outer scale value is the median of the remaining ${\cal L}_{0,i}$ after filtering. 

%
%
\subsection{Error analysis}
\label{par:errorsL0}
The estimation of $\lo$ is made from the ratios $Q_{i}$ by inverting Eq.~\ref{eq:ratioQL0}. The error $\delta_Q$ comes from errors on variances which propagate to $Q_{i}$ via Eq.~\ref{eq:ratioQ}. 

To increase  accuracy, we perform a rolling average of measured variances (they are calculated every 2mn) over time intervals of $T$ (set to $T=$10 minutes), corresponding to an average of $N_v=5$ individual variances, thus reducing the error by $\sqrt{N_v}$ on variances. The relative error  $\delta_Q$ on $Q_i$ expresses as
\be
\frac{\delta_Q}{Q_i}=\frac 1{\sqrt{N_v}} \left(\frac{\delta\sigma_{Di}^2}{\sigma_{Di}^2}+ \frac{\delta\sigma_{l|t}^2}{\sigma_{l|t}^2}\right)
\ee
Taking only the statistical error on variances (they dominate indeed, as discussed in Sect.~\ref{par:staterr}), we obtain $\frac{\delta_Q}{Q_i}\simeq 5\%$ for $N=1024$ images and $N_v=5$. The error $\delta \lo$ on the outer scale is obtained by the finite difference
\be
\delta \lo=\lo'(Q_i)\ \delta_Q
\ee
where the derivative $\lo'(Q_i)$ is calculated from  Eq.~\ref{eq:ratioQL0}.
\begin{figure}
\begin{center}
\includegraphics[width=8cm]{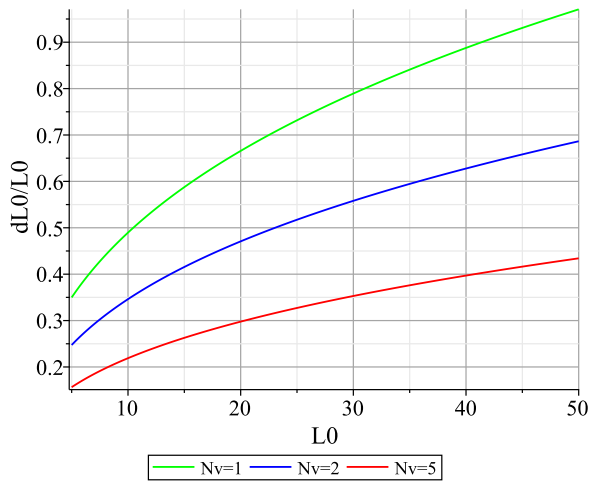}
\caption{Relative error on $\lo$ as a function of $\lo$ for different values of $N_v$ (number of averaged variances). The number of images in a sequence is $N=1024$.}
   \label{fig:errL0}
\end{center}
\end{figure}
The expected relative error $\frac{\delta\lo}{\lo}$ (due to the statistical error) is shown in Fig.\ref{fig:errL0}. Three curves are plotted for different values of $N_v$ in the range $\lo\in[5,50]$m. Both show that low $\lo$ values are estimated with better precision. With $N_v=1$ (no variance averaging) is impossible to obtain reliable values of $\lo$ (relative error is $\sim$70\% for $\lo=20$m). An average of at least $N_v=5$ individual variances is necessary to obtain acceptable error bars ($\frac{\delta\lo}{\lo}\simeq 30$\%  for $\lo=20$m). The drawback is that one obtains estimations of $\lo$ smoothed over time intervals 10mn with $N_v=5$. This is greater than the characteristic time of outer scale fluctuations, whose value, estimated by GSM, is of the order of 6mn \cite{Ziad16}. 

{
The statistical error is not the only contribution to the total uncertainty, especially for absolute variances which are contaminated by vibrations. A measure on their effect on $\lo$ can be made from the remaining distribution of $H$ invariants after filtering (see Section~\ref{par:dataprocL0}). The thresholds on $H_{l|t}$ to filter the data were obtained as a trade-off between data quality and the number of variances kept for outer scale estimation. The remaining $H$ distribution has a dispersion $\Delta H\simeq 0.1$ around the nominal value. This results into an error $\Delta\lo$ on the outer scale. To estimate it, we rewrite Eq.~\ref{eq:Hinv} as
\be
H_{l|t}=Q_{i,l|t}-Q_{3,l|t}
\ee
so that 
\be
\Delta H\simeq \Delta Q_{i,l|t}+\Delta Q_{3,l|t} \simeq 0.1
\ee
corresponding to an uncertainty $\Delta Q_{i,l|t}\simeq 0.05$ on the ratios $Q$. Writing
\be
\Delta Q_{i,l|t}=\frac{\partial Q_{i,l|t}}{\partial\lo} \: \Delta\lo
\ee
and making use of Eq.~\ref{eq:ratioQL0} to calculate $\frac{\partial Q_{i,l|t}}{\partial\lo}$, we found that the resulting relative error on $\lo$ is of the order of 50\% for $\lo$ around 20m.

We are currently working on improvements on the algorithm of $\lo$ calculation to find better metrics and to reduce the effect of vibrations, an issue on small telescopes.
}

%
%

\begin{figure*}
\begin{center}
\includegraphics[width=\textwidth]{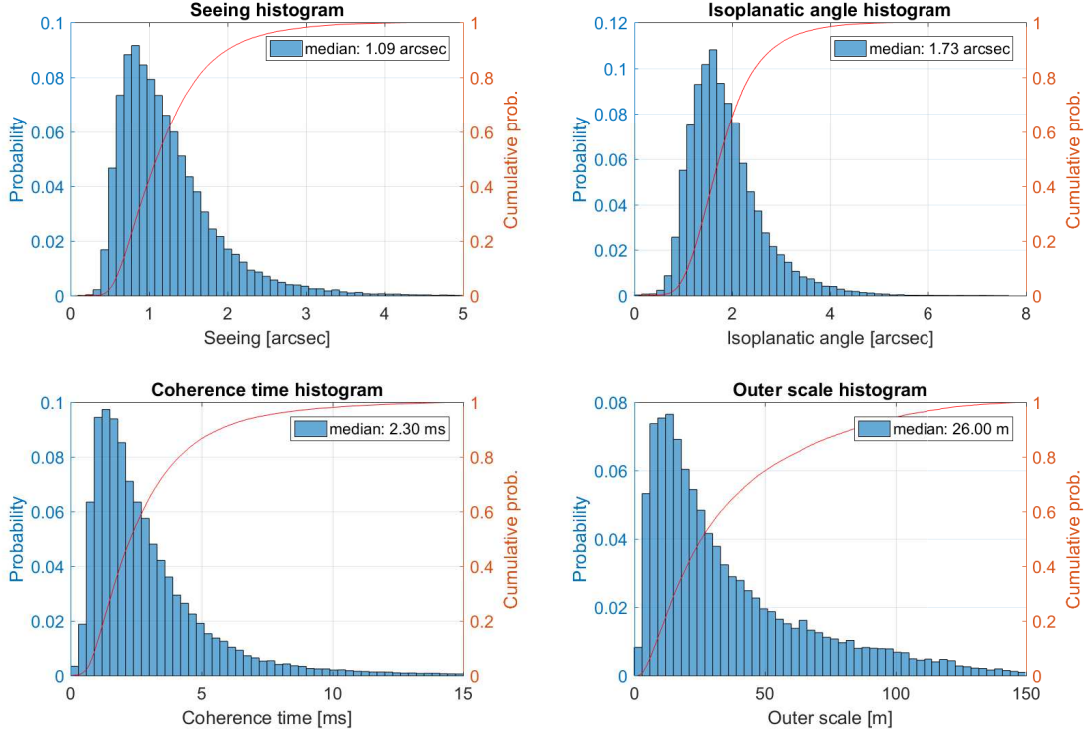}
\caption{Histograms of turbulence parameters at Calern, calculated at the wavelength $\lambda=0.5\mu$m.}
   \label{fig:paramshisto}
\end{center}
\end{figure*}

\section{First long-term GDIMM statistics}
\label{par:results}
\begin{table}
\begin{tabular}{l|c|c|c|c|c|c}
          & $\epsilon_0$  &  $\theta_0$  & $\tau_0$  &  $\lo$  & $\bar h$   & $\bar v$   \\
					&  [$''$] &   [$''$] &  [ms] &   [m] &  [m] &   [m/s] \\ \hline
Median    				& 1.09  & 1.73  & 2.30  & 26.00 & 3436 & 12.84\\
Mean      				& 1.23  & 1.86  & 3.10  & 37.14 & 3698 & 13.59 \\
Std. dev.					 & 0.52  & 0.65  & 1.80  & 29.20 & 1566 & 5.47\\
$1^{st}$ quartile & 0.80  & 1.35  & 1.40  & 13.50 & 2504 & 9.24 \\
$3^{rd}$ quartile & 1.49  & 2.21  & 3.80  & 51.00 & 4618 & 16.74\\ 
$1^{st}$ centile & 0.45  & 0.58  & 0.50  & 3.10 & 1121 & 3.03\\
Last     centile & 3.37  & 4.27  & 14.90  & 142.25 & 10279 & 30.23\\ \hline
Paranal   			& 0.81 & 2.45 & 2.24 & 22 & 3256 & 17.3 \\
La Silla  			 & 1.64  &  1.25 & 1.46  & 25.5 & 3152  & 13.1  \\
Mauna Kea   		& 0.75 & 2.94 & 2.43 & 24 & 2931 &  17.2 \\ \hline
\end{tabular}
\caption{Statistics of turbulence parameters measured at Calern (at the wavelength $\lambda=0.5\mu$m) during the period June 2015--October 2018. Paranal, La Silla and Mauna Kea values are from the GSM database.}
   \label{table:paramstat}
	\end{table}
A total of 70097 turbulence parameter measurements (22698 for ${\cal L}_0$) were collected at Calern observatory during the 3$\frac 1 2$~year period from June~2015 to October~2018. Half of the data were obtained during the Summer season (June to September) where meteo conditions are better. Statistics are presented in Table~\ref{table:paramstat} for the 4 turbulence parameters ($\epsilon_0$, $\theta_0$ $\tau_0$, $\lo$) and for the equivalent turbulence altitude (Eq.~\ref{eq:hmoy}) and the effective wind speed (Eq.~\ref{eq:tau0}). Histograms are displayed in Fig.~\ref{fig:paramshisto} and show a classical log-normal shape for all parameters. Compared to other astronomical sites in the world (examples for Paranal, La Silla and Mauna Kea are given in Table~\ref{table:paramstat}) show that the Calern plateau is an average site.

The seeing is slightly lower in summer, we measured a median value of $0.96''$ in July and August (the median winter seeing during the period November--January is 1.21$''$). As a consequence, the median coherence time is higher in summer (3.2ms in July--August, 2.40ms in November--January). The outer scale $\lo$ has values similar to other sites such as Mauna Kea or La Silla.

\begin{figure*}
\begin{center}
\includegraphics[width=8cm]{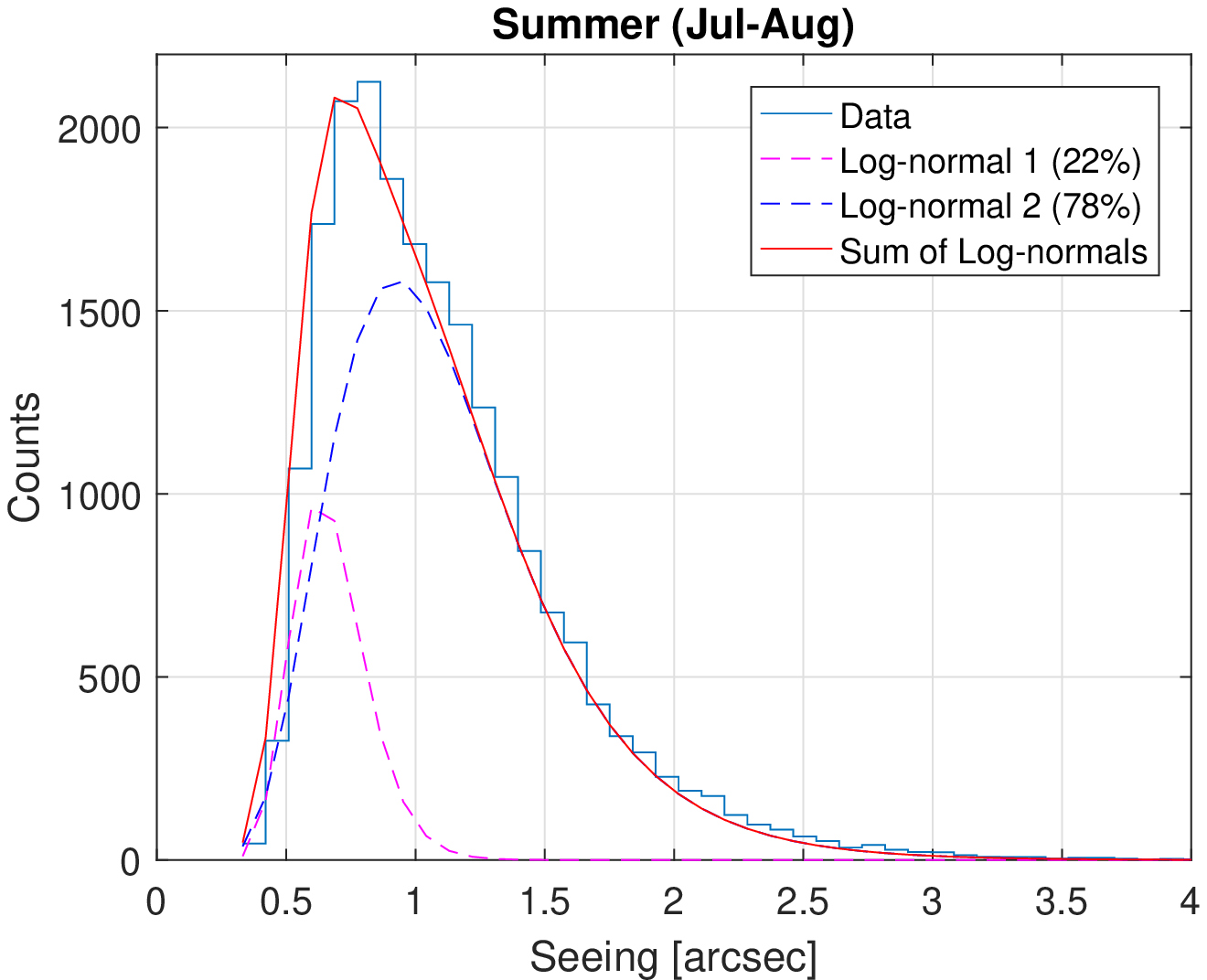}
\includegraphics[width=8cm]{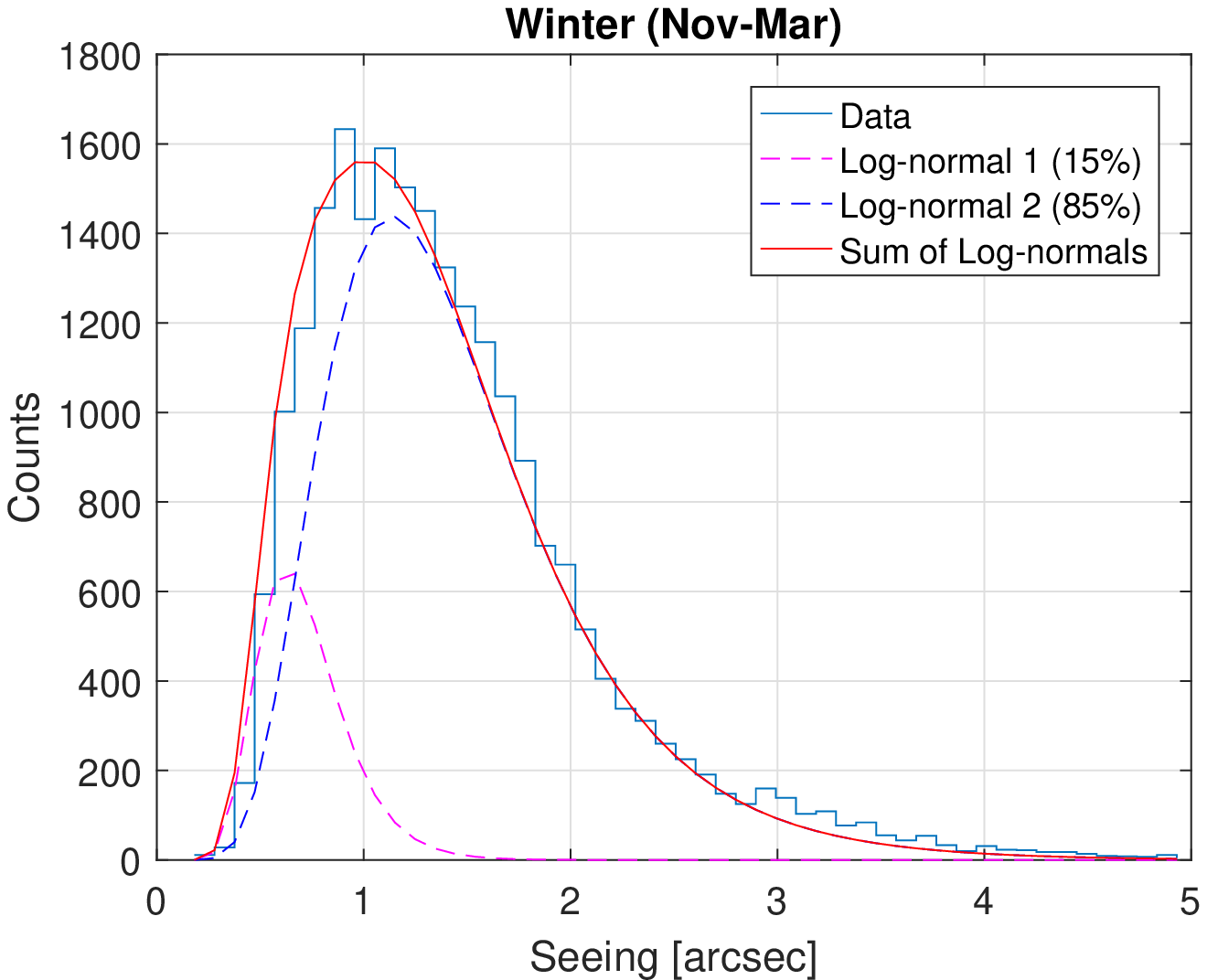}
\caption{Left: seeing histogram for the summer (July--August). Right: seeing histogram for the winter (Nov--March). Superimposed curves are a least-square fit by a sum of two log-normal distributions (individual log-normal curves are dashed lines). The percentages corresponding to each log-normal in indicated in the legend.}
   \label{fig:histseeingsummerfit}
\end{center}
\end{figure*}

Sequences of several hours of good seeing were sometimes observed, which is a good point for this site (and already known by ``old'' observers on interferometers during the 80's and 90's).Fig.~\ref{fig:histseeingsummerfit} displays seasonal seeing histograms, calculated for the summer (July and August) and the winter (November--March). They appear to be well modelled by a sum of two log-normal functions (they appear as dashed curves on the plots, their sum is the solid line). This is an evidence of the existence of two regimes: a ``good seeing'' distribution with a median value $\epsilon_1$ and a ``medium seeing'' situation with a median value $\epsilon_2$. In summer, we have $\epsilon_1=0.63''$ (the good seeing distribution contains 22\% of the data) and $\epsilon_2=0.95''$ (78\% of the data). In winter we have $\epsilon_1=0.66''$ (15\% of the data) and  $\epsilon_2=1.15''$ (85\% of the data). 

The equivalent turbulence altitude $\bar h$ has a median value around 3km, which is comparable to other classical sites. However we noticed a difference between the summer and the winter. During the 2 months of July and August, the median value of $\bar h$ was 3940m, while it is only 2870m in winter (November to March). Situations with a high value of $\bar h$ correspond to less turbulence in the ground layer, giving good seeing conditions as the ground layer is the main contributor to the total seeing.

\begin{figure*}
\begin{center}
\includegraphics[width=8cm]{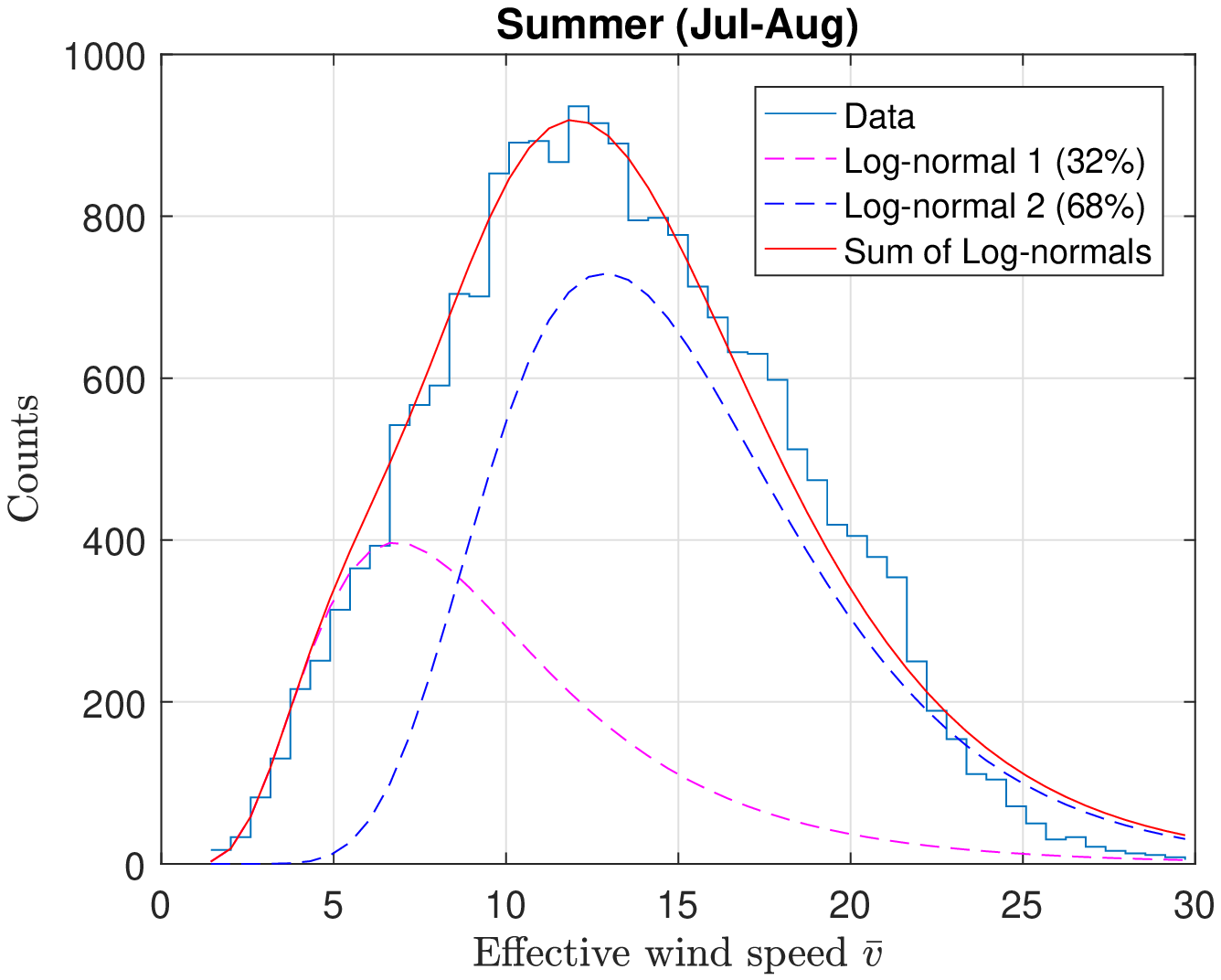}
\includegraphics[width=8cm]{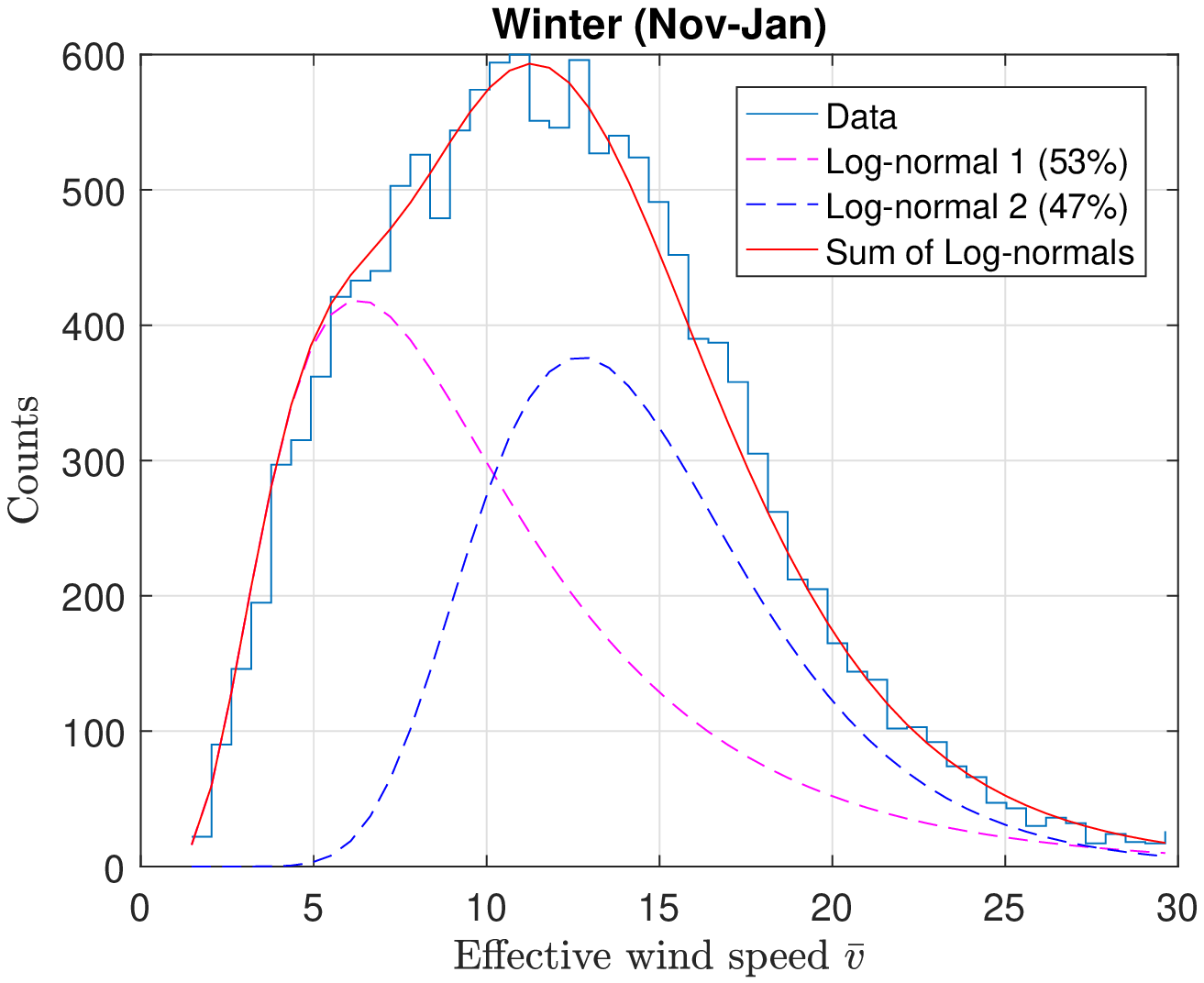}
\caption{Histograms of the effective wind speed in winter (November--January). Superimposed curves are a least-square fit by a sum of two log-normal distributions. Left: summer histogram. Right: winter histogram.}
   \label{fig:histoveffsum}
\end{center}
\end{figure*}

As for the seeing, the effective wind speed histograms (Fig.~\ref{fig:histoveffsum}) are bimodal and can be modelled by the sum of two log-normal functions. They peak at $\bar v_1=6.7$m/s and $\bar v_2=13$m/s both for the summer and the winter. They contain respectively 32\% and 68\% of the data in summer, these proportions go to 53\% and 47\% in winter. The value $\bar v_1=6.7$m is indeed close to the median ground wind speed $v_G=5.7$m/s measured by the meteo station. 

%
%
\section{Conclusions}
\label{par:conclusion}

We have presented the GDIMM, a new turbulence monitor aiming at measuring the 4 integrated parameters
of the optical turbulence. GDIMM is a small instrument, easy to transport to make measurements at any site in the world, and was designed to provide a monitoring of the four integrated parameters of the atmospheric turbulence, i.e. seeing, isoplanatic angle, coherence time and outer scale. 

Seeing measurements are given by differential motion, according to a well established theory and to an instrumental concept that makes them robust to telescope vibrations \cite{Sarazinroddier90, Verninmunoz95}. Isoplanatic angle measurements are made via the scintillation, following here again a well-known technique \cite{Looshogge79}, which has become popular thanks to its simplicity. It appears to give satisfactory results when compared to other techniques \cite{Ziad18b}. We indeed used intensively these two techniques to measure $\epsilon_0$ and $\theta_0$ during the campaigns of site testing of the site of Dome~C in Antarctica (see \cite{Aristidi12} and references therein).

The method for estimating the coherence time from the decorrelation time of AA fluctuations is recent. It was proposed a few years ago \cite{Ziad12} and is based upon analytical developments by \cite{Conan00}. First tests on reprocessed GSM data and comparisons with radiosoundings \cite{Ziad12} showed the pertinence of the method. The instrumental concept is simple, compared to other monitors such as the MASS-DIMM \cite{Kornilov07}, the only requirement is to have a  camera allowing a high framerate (at least 100 frames per second) to properly sample the AA decorrelation time. After GSM in the past, the GDIMM is now, to our knowledge, the first monitor to { use this method routinely to calculate $\tau_0$.
}
A true asset of GDIMM is the possibility to measure the outer scale. In particular, obtaining reliable values of ${\cal L}_0$ is a challenge with small instruments, and this parameter is often neglected, though it has a strong impact of high angular resolution techniques, especially for extremely large telescopes (see the recent review by \cite{Ziad16}). We proposed here a method based on the ratios of absolute to differential motions. It is simple and can work with any DIMM or Shack-Hartmann based monitor, but requires good stability of the telescope mount since it is sensitive to vibrations.



A portable version of the GDIMM has been developped in parallel to the Calern one, to perform turbulence measurements at any site on the world. Discussions with the ESO (European Southern Observatory) are currently in progress to make GDIMM and PML observations at Paranal and compare with the ESO Astronomical Site Monitor \cite{Chiozzi16}.

\section{Acknowledgments}
%
%
We would like thank Jean-Marie Torre and Herv\'e Viot, from the Calern technical staff, for their valuable help on the electronics of the instrument. Thanks also to M. Marjani who worked on our data during his master thesis. The CATS project has been done under the financial support of CNES, Observatoire de la C\^ote d'Azur, Labex First TF, AS-GRAM, Federation Doblin, Universit\'e de Nice-Sophia Antipolis and R\'egion Provence Alpes
C\^ote d'Azur.

\bibliography{biblio}   

\begin{thebibliography}{10}

\bibitem{Labeyrie70}
{Labeyrie}, A., ``{Attainment of Diffraction Limited Resolution in Large
  Telescopes by Fourier Analysing Speckle Patterns in Star Images},'' {\em
  A\&A}~{\bf 6},  85 (May 1970).

\bibitem{Labeyrie75}
{Labeyrie}, A., ``{Interference fringes obtained on VEGA with two optical
  telescopes},'' {\em ApJL}~{\bf 196},  L71--L75 (Mar. 1975).

\bibitem{Rousset90}
{Rousset}, G., {Fontanella}, J.~C., {Kern}, P., {Gigan}, P., and {Rigaut}, F.,
  ``{First diffraction-limited astronomical images with adaptive optics},''
  {\em A\&A}~{\bf 230},  L29--L32 (Apr. 1990).

\bibitem{Carbillet17}
{Carbillet}, M., {Aristidi}, E., {Giordano}, C., and {Vernin}, J. {\em
  MNRAS}~{\bf 471},  3043 (2017).

\bibitem{Winker91}
{Winker}, D.~M., ``{Effect of a finite outer scale on the Zernike decomposition
  of atmospheric optical turbulence.},'' {\em JOSA A}~{\bf 8},  1568--1573
  (1991).

\bibitem{Ziad00}
Ziad, A., Conan, R., Tokovinin, A., Martin, F., and Borgnino, J. {\em Appl.
  Opt.}~{\bf 39},  5415 (2000).

\bibitem{Aristidi14}
{Aristidi}, E., {Fante{\"i}-Caujolle}, Y., {Ziad}, A., {Dimur}, C.,
  {Chab{\'e}}, J., and {Roland}, B., ``{A new generalized differential image
  motion monitor},'' in [{\em Ground-based and Airborne Telescopes
  V}{\nolinebreak\hspace{0.1em}]},  {\em SPIE proc.} {\bf 9145},  91453G (July
  2014).

\bibitem{Sarazinroddier90}
Sarazin, M. and Roddier, F. {\em A\&\/A}~{\bf 227},  294 (1990).

\bibitem{Samain08}
{Samain}, E., {Abchiche}, A., {Albanese}, D., {Geyskens}, N., {Buchholtz}, G.,
  {Drean}, A., {Dufour}, J., {Eysseric}, J., {Exertier}, P., {Pierron}, F.,
  {Pierron}, M., {Martinot}, G., L., {Paris}, J., {Torre}, J.-M., and {Viot},
  H., ``{MEO : The New French Lunar Laser Ranging Station},'' in [{\em 16th
  International Workshop on Laser Ranging}{\nolinebreak\hspace{0.1em}]},   88
  (Oct. 2008).

\bibitem{Bendjoya12}
{Bendjoya}, P., {Abe}, L., {Rivet}, J.-P., {Su{\'a}rez}, O., {Vernet}, D., and
  {M{\'e}karnia}, D., ``{C2PU: An original mix of research and pedagogy at
  Observatoire de la C{\^o}te d'Azur},'' in [{\em proc. of the
  SF2A-2012}{\nolinebreak\hspace{0.1em}]},   643--648 (Dec. 2012).

\bibitem{Tokovinin02}
Tokovinin, A. {\em Pub. Astron. Soc. Pacific}~{\bf 114},  1156 (2002).

\bibitem{Ziad17}
Ziad, A. et~al. {\em AO4ELT 5, Tenerife, Spain, June 25-30} (2017).

\bibitem{Ziad18}
{Ziad}, A., {Chab{\'e}}, J., {Fantei-Caujolle}, Y., {Aristidi}, E., {Renaud},
  C., and {Ben Rahhal}, M., ``{CATS: an autonomous station for atmospheric
  turbulence characterization},'' in [{\em SPIE Conference
  Series}{\nolinebreak\hspace{0.1em}]},  {\em SPIE Conference Series} {\bf
  10703},  107036L (July 2018).

\bibitem{Aristidi18}
{Aristidi}, E., {Fant\'ei-Caujolle}, Y., {Chab{\'e}}, J., {Renaud}, C., {Ziad},
  A., and {Ben Rahhal}, M., ``{Turbulence monitoring at the Plateau de Calern
  with the GDIMM instrument},'' in [{\em SPIE Conference
  Series}{\nolinebreak\hspace{0.1em}]},  {\em SPIE Conference Series} {\bf
  10703},  107036U (July 2018).

\bibitem{Ziad16}
{Ziad}, A., ``{Review of the outer scale of the atmospheric turbulence},'' in
  [{\em SPIE Conference on Adaptive Optics Systems
  V}{\nolinebreak\hspace{0.1em}]},  {\em SPIE Conference Series} {\bf 9909},
  99091K (July 2016).

\bibitem{Frieden83}
{Frieden}, B.~R.,  [{\em {Probability, statistical optics, and data
  testing.}}{\nolinebreak\hspace{0.1em}]}, Berlin: Springer, 1983 (1983).

\bibitem{Ziad94}
Ziad, A., Borgnino, J., Martin, F., and Agabi, A. {\em A\&\/A}~{\bf 282},  1021
  (1994).

\bibitem{Looshogge79}
Loos, G. and Hogge, C. {\em Appl. Opt.}~{\bf 18},  15 (1979).

\bibitem{Roddier82}
{Roddier}, F., {Gilli}, J.~M., and {Vernin}, J. {\em Journal of Optics}~{\bf
  13},  63--70 (1982).

\bibitem{Roddier81}
Roddier, F. {\em Progress in Optics}~{\bf 19},  281 (1981).

\bibitem{Ziad12}
{Ziad}, A., {Borgnino}, J., {Dali Ali}, W., {Berdja}, A., {Maire}, J., and
  {Martin}, F., ``{Temporal characterization of atmospheric turbulence with the
  Generalized Seeing Monitor instrument},'' {\em Journal of Optics}~{\bf 14},
  045705 (Apr. 2012).

\bibitem{Conan00}
{Conan}, R., {Borgnino}, J., {Ziad}, A., and {Martin}, F., ``{Analytical
  solution for the covariance and for the decorrelation time of the angle of
  arrival of a wave front corrugated by atmospheric turbulence},'' {\em Journal
  of the Optical Society of America A}~{\bf 17},  1807--1818 (Oct. 2000).

\bibitem{ZiadThese}
{Ziad}, A., {\em {Estimation des échelles limites de cohérence spatiale des
  fronts d'onde et optimisation des observations a haute résolution angulaire
  en astronomie}}, PhD thesis, Universit\'e de Nice -- Sophia Antipolis, France
  (1993).

\bibitem{Verninmunoz95}
Vernin, J. and Munoz-Tunon, C. {\em Pub. Astron. Soc. Pacific}~{\bf 107},  265
  (1995).

\bibitem{Ziad18b}
{Ziad}, A., {Aristidi}, E., {Chab\'e}, J., and {Borgnino}, J., ``{On the
  isoplanatic patch size in High Angular Resolution Techniques},'' {\em MNRAS,
  submitted}  (2018).

\bibitem{Aristidi12}
{Aristidi}, E., ``{Dome C site testing: Long term statistics of integrated
  optical turbulence parameters at ground level},'' in [{\em Proc.
  SF2A-2012}{\nolinebreak\hspace{0.1em}]},  {Boissier}, S., {de Laverny}, P.,
  {Nardetto}, N., {Samadi}, R., {Valls-Gabaud}, D., and {Wozniak}, H., eds.,
  697--701 (Dec. 2012).

\bibitem{Kornilov07}
{Kornilov}, V., {Tokovinin}, A., {Shatsky}, N., {Voziakova}, O., {Potanin}, S.,
  and {Safonov}, B., ``{Combined MASS-DIMM instruments for atmospheric
  turbulence studies},'' {\em MNRAS}~{\bf 382},  1268--1278 (Dec. 2007).

\bibitem{Chiozzi16}
{Chiozzi}, G., {Sommer}, H., {Sarazin}, M., {Bierwirth}, T., {Dorigo}, D.,
  {Vera Sequeiros}, I., {Navarrete}, J., and {Del Valle}, D., ``{The ESO
  astronomical site monitor upgrade},'' in [{\em Software and
  Cyberinfrastructure for Astronomy IV}{\nolinebreak\hspace{0.1em}]},  {\em
  SPIE proc.} {\bf 9913},  991314 (Aug. 2016).

\end{thebibliography}
\bibliographystyle{spiebib}   



\end{document}